\documentclass[fleqn,usenatbib]{mnras}
\usepackage[T1]{fontenc}
\usepackage{ae,aecompl}
\usepackage[dvipdfmx]{graphicx}
\usepackage{multirow}
\usepackage{amsmath}
\usepackage{amssymb}
\usepackage{lscape}
\usepackage{ulem,color}

\newcommand{\Msun}{\ensuremath{\mathrm{M}_\odot}}
\newcommand{\MP}{\ensuremath{m_{\rm{p}}}}
\newcommand{\ME}{\ensuremath{m_{\rm{e}}}}

\newcommand{\Mej}{\ensuremath{{M}_{\rm{ej}}}}
\newcommand{\Ekin}{\ensuremath{{E}_{\rm{kin}}}}

\newcommand{\epe}{\ensuremath{\varepsilon_{\rm{e}}}}
\newcommand{\bepe}{\ensuremath{\bar{\varepsilon}_{\rm{e}}}}
\newcommand{\bepen}{\ensuremath{\bar{\varepsilon}_{\rm{e,-1}}}}
\newcommand{\epb}{\ensuremath{\varepsilon_{\rm{B}}}}

\newcommand{\SGM}{\ensuremath{\sigma_{\rm{T}}}}

\newcommand{\gamm}{\ensuremath{{\gamma}_{\rm{m}}}}
\newcommand{\num}{\ensuremath{{\nu}_{\rm{m}}}}
\newcommand{\nua}{\ensuremath{{\nu}_{\rm{a}}}}

\title[Radio constraints on TDEs]{Radio constraint on outflows from tidal disruption events}

\author[Matsumoto \& Piran]{Tatsuya Matsumoto,$^{1,2,3}$\thanks{E-mail: tatsuya.matsumoto@mail.huji.ac.il}\thanks{JSPS Research Fellow} and Tsvi Piran$^{1}$\thanks{E-mail: tsvi.piran@mail.huji.ac.il}
\\
$^{1}$Racah Institute of Physics, Hebrew University, Jerusalem, 91904, Israel\\
$^{2}$Research Center for the Early Universe, Graduate School of Science, University of Tokyo, Tokyo 113-0033, Japan\\
$^{3}$Department of Physics, Graduate School of Science, University of Tokyo, Tokyo 113-0033, Japan\\
}

\pubyear{2020}

\begin{document}
\label{firstpage}
\pagerange{\pageref{firstpage}--\pageref{lastpage}}
\maketitle

\begin{abstract}
Radio flares from tidal disruption events (TDEs) are generally interpreted as synchrotron emission arising from the interaction of an outflow with the surrounding circumnuclear medium (CNM). 
We generalize the common equipartition analysis to be applicable in cases lacking a clear spectral peak or even with just an upper limit.
We show that, for detected events, there is a lower limit on the combination of the outflow's velocity $v$ and solid angle $\Omega$, $\simeq v\Omega^{a}$ (with $a \simeq 0.5$).
Considering several possible outflow components accompanying TDEs, we find that: Isotropic outflows such as disk winds with $v\sim10^4\,\rm km\,s^{-1}$ and $\Omega = 4 \pi$ can easily produces the observed flares; The bow shock of the unbound debris has a wedge-like geometry and it must be geometrically thick with $\Omega\gtrsim1$. A fraction of its mass ($\gtrsim 0.01 \Msun$) has to move at $v \gtrsim 2 \times 10^4\,\rm km\,s^{-1}$; Conical Newtonian outflows such as  jets can also be a radio source but both their velocity and the CNM density should be larger than those of isotropic winds by a factor of $\sim(\Omega/4\pi)^{-0.5}$. 
Our limits on the CNM densities are typically 30-100 times larger than those found by previous analysis that ignored non-relativistic electrons.
We also find that late (a few years after the TDE) radio upper-limits rule out energetic, $\sim 10^{51-52}\,\rm erg$, relativistic jets like the one observed in TDE Sw J1644+57, implying that such jets are rare.
\end{abstract}

\begin{keywords}
transients: tidal disruption events
\end{keywords} 

\section{Introduction}
A star that approaches a supermassive black hole (BH) close enough will be torn apart leading to a tidal disruption event (TDE) \citep{Hills1975,Rees1988}.
After the disruption, about half of the stellar debris is bound and falls back to the BH.
If the debris forms an accretion disk rapidly, we observe the event as a bright X-ray flare at the galactic center.
Actually, the first events considered to be TDEs were discovered in the X-ray band \citep[see][for reviews]{Komossa2015,Saxton+2020}.
Recently, more TDEs have been detected in optical/UV bands \citep{vanVelzen+2020}.

\begin{figure}
\begin{center}
\includegraphics[width=85mm, angle=0]{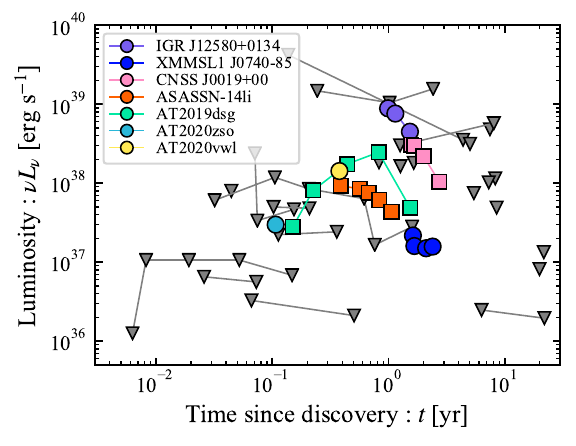}
\caption{Radio light curves and upper limits of TDEs. The colored data points are detected events at frequency of $\nu\simeq5\,\rm GHz$ and gray down triangles are upper limits. The squares (circles) mean the observations at which the spectral peak is (not) detected. On-axis jetted TDEs, Sw J1644+57, Sw J2058+05, and Sw J1112-82, whose radio luminosity is much larger $\sim10^{40-42}\,\rm erg\,s^{-1}$, are not shown (see figure 1 in \citealt{Alexander+2020} for these events).}
\label{fig lc}
\end{center}
\end{figure}

Some TDEs also produce radio flares \citep[see][for a review]{Alexander+2020}.
The first-discovered radio emission was from a peculiar TDE, Sw J1644+57 (hereafter Sw1644), which launched a relativistic jet \citep{Bloom+2011,Burrows+2011,Levan+2011,Zauderer+2011}.
While jetted TDEs make very bright radio flares $L\sim10^{40-42}\,\rm erg\,s^{-1}$, their fraction of the whole TDE population is small.
On the other hand, radio emissions have been observed also from optical/UV TDEs.
The prototype of those is the radio flare of the optical TDE, ASASSN14-li \citep{Alexander+2016,vanVelzen+2016}.
This flare was detected $\sim100\,\rm days$ after the discovery in the optical band and its luminosity is $\sim10^3$ times smaller than those of jetted TDEs.
Some optical TDEs show similar radio flares to that of ASASSN-14li as shown in Fig. \ref{fig lc}.

A natural interpretation of the radio emission is that it arises from the interaction of an outflow launched by the TDE with the circumnuclear medium (CNM) surrounding the BH.
This produces a blast wave and at the shock front, the magnetic field is amplified and electrons are accelerated to a relativistic energy, which produces synchrotron emission. 
Therefore, the radio detection and even upper limits are useful to constrain the outflow properties and CNM density around galactic centers.

The origin of outflows causing the radio flares is still debated  while the number of radio TDEs increases and we have more data to address this question.
Several channels can launch outflows from TDEs and each one of them can potentially produce the observed radio.
An unavoidable outflow is the unbound stellar debris which is launched at the moment of disruption \citep{Krolik+2016,Yalinewich+2019b}.
While it is confined to the stellar orbital plane, a significant mass $\simeq0.5\,\Msun$ is ejected at a high velocity $\sim10^4\,\rm km\,s^{-1}$.
The second potential source arises if a compact accretion disk forms by the infalling bound stellar material.
Such a disk accretes at super-Eddington rate and can launch a strong outflow \citep{Strubbe&Quataert2009,Metzger&Stone2016}.
The third possibility involves relativistic jets that have been detected in some TDEs and can also produce radio emissions \citep{Giannios&Metzger2011}.

In this work, we analyze the currently observed radio TDEs\footnote{Recently \cite{Horesh+2021} reported detection of multiple radio flares for ASASSN-15oi. We defer analyzing this event to a future work.} as well as all currently available  radio upper-limits and infer the outflow properties and CNM density for different outflow models.
So far radio TDEs have been analyzed by the equipartition method \citep{Chevalier1998,BarniolDuran+2013} that can be used when the spectral peak is observed \citep{BarniolDuran&Piran2013,Zauderer+2013,Alexander+2016,Krolik+2016,Eftekhari+2018,Anderson+2020,Stein+2021,Cendes+2021,Cendes+2021b} or by both analytical and numerical modeling for bright events such as Sw1644 \citep{Berger+2012,Metzger+2012,Mimica+2015}.
We develop a general framework that enables us to constrain properties of the outflow and the surrounding matter from more limited radio data  (e.g. without observation of the spectral peak) and even in cases where only upper limits are available.

The paper is organized as follows.
In \S \ref{sec method}, we formulate our method describing synchrotron formulae and dynamics of outflows.
In \S \ref{sec app} we apply the method to different sources of outflows and derive constraints on the outflow's properties and density profile from the observed radio data.
We consider spherical outflow resulting from super-Eddington winds (\S \ref{sec wind}), wedge-shape unbound debris (\S \ref{sec unbound}), and conical outflow (\S \ref{sec conical}) corresponding to Newtonian jets.
We discuss the late Newtonian phase of relativistic jets in \S \ref{sec jet} and obtain limits on the allowed jet energy.
We summarize and discuss the implications of our results in \S \ref{sec summary}.

\section{Method}\label{sec method}
\subsection{Synchrotron emission}\label{sec emission}
We describe the method to calculate the synchrotron flux based on \cite{Piran+2013,Ricci+2021}.
The CNM surrounding the BH in the TDEs is much denser than the interstellar medium around short gamma-ray bursts ($\lesssim1\,\rm cm^{-3}$) and the synchrotron self-absorption (SSA) effect becomes important which shapes the observed spectrum.
As most detected TDEs are at relatively small redshifts, we neglect the redshift effect in the following equations which can be easily restored.
We consider an outflow traveling at a velocity $v$ in the CNM with the number density of $n$ at the shock front.
The amplified magnetic field is given by an argument in which a fraction $\varepsilon_{\rm B}$ of the post-shock thermal energy is transferred to the magnetic field energy:
\begin{align}
B&=(8\pi\epb \MP  nv^2)^{1/2}
	\label{eq B}\\
&\simeq6.5\times10^{-4}{\,\rm G\,}\varepsilon_{\rm B,-2}^{1/2}n_0^{1/2}v_9,
	\nonumber
\end{align}
where $\MP$ is the proton mass.
We use the notation $Q_x=Q/10^x$ in cgs units unless otherwise specified.
A fraction $\varepsilon_{\rm e}$ of energy is also used to accelerate relativistic electrons in a power-law distribution.
The minimum Lorentz factor of electrons and the corresponding synchrotron frequency are given by
\begin{align}
\gamm&={\rm max}\biggl[2,\,\frac{\MP}{4\ME c^2}\bepe v^2\biggl]\simeq{\rm max}\big[2,\,0.051\,\bepen v_{9}^2\big],\\
\num&=\gamm^2\frac{eB}{2\pi\ME c}
	\label{eq nu_m}\\
&\simeq\begin{cases}
7.2\times10^{3}{\,\rm Hz\,}\,\varepsilon_{\rm B,-2}^{1/2}n_0^{1/2}v_9 &:\,v<v_{\rm DN},\\
4.7{\,\rm Hz\,}\,\bepen^2\varepsilon_{\rm B,-2}^{1/2}n_0^{1/2}v_{9}^5 &:\,v_{\rm DN}<v,
\end{cases}
	\nonumber
\end{align}
respectively, where $\ME$ is the electron mass, $c$ is the speed of light, $e$ is the elementary charge, and we define $\bepe\equiv4\epe(p-2)/(p-1)$ with the electron distribution's power-law index $p$.
When the outflow's velocity is lower than the critical value
\begin{align}
v<v_{\rm DN}=\biggl(\frac{8\ME}{\MP\bepe}\biggl)^{1/2}c\simeq6.3\times10^{4}{\,\rm km \,s^{-1}\,}\bepen^{-1/2},
\end{align}
the Lorentz factor is fixed to $\gamm=2$.
We call this regime as the deep-Newtonian phase \citep{Huang&Cheng2003,Sironi&Giannios2013}.
Hereafter we normalize the velocity by $10^{9}\,\rm cm\,s^{-1}$ regardless of each phase while an outflow with this velocity is in the deep-Newtonian phase.

The spectral power from an electron with the Lorentz factor $\gamm$ is given by
\begin{align}
P_{\num}&\simeq\frac{\frac{4}{3}\SGM c \gamm^2\frac{B^2}{8\pi}}{\num}
	\label{eq p nu_m}\\
&\simeq2.5\times10^{-25}{\rm erg\,s^{-1}\,Hz^{-1}\,}\varepsilon_{\rm B,-2}^{1/2}n_0^{1/2}v_9,
    \nonumber
\end{align}
where $\sigma_{\rm T}$ is the Thomson cross section.

We estimate the number of electrons by
\begin{align}
N_{\rm e}\simeq\Omega nR^3,
	\label{eq electron number}
\end{align}
where $\Omega$ is the solid angle subtended by the outflow.
Examination of the Milky Way galactic center \citep{Baganoff+2003,Gillessen+2019} as well as the analyses of radio TDEs \citep[e.g.][]{Alexander+2016,Krolik+2016} suggest that the CNM density in galactic nuclear regions has a power-law like profile, $n\propto R^{-k}$ ($k<3$), where $R$ is the distance from the BH.
As we are not considering here the light curves but only the emission at some given moments of time we do not specify the density profile in this work but use instead only the density at the shock radius.
As long as the density profile is shallow enough $k<3$, this estimate is accurate up to a numerical factor, $1/(3-k)$.

Noting that the number of radiating electrons is reduced by a factor of $(v/v_{\rm DN})^2$ in the deep-Newtonian phase, we calculate the flux density at $\num$
\begin{align}
F_{\num}&=\frac{N_{\rm e}\min\big[\big(\frac{v}{v_{\rm DN}}\big)^2,\,1\big]P_{\num}}{4\pi d_{\rm L}^2}
	\label{eq f nu_m}\\
&\simeq\begin{cases}
0.63{\,\rm \mu Jy\,}\bepen\varepsilon_{\rm B,-2}^{1/2}n_0^{3/2}v_9^3R_{17}^3\big(\frac{\Omega}{4\pi}\big)d_{\rm L,27}^{-2}&:\,v<v_{\rm DN},\\
25{\,\rm \mu Jy\,}\varepsilon_{\rm B,-2}^{1/2}n_0^{3/2}v_{9}R_{17}^3\big(\frac{\Omega}{4\pi}\big)d_{\rm L,27}^{-2}&:\,v_{\rm DN}<v,
\end{cases}
	\nonumber
\end{align}
where $N_{\rm e}$ and $d_{\rm L}$ are the total number of electrons and the luminosity distance, respectively.

The SSA frequency, which is typically larger than $\num$ for our parameters, is given by 
\begin{align}
\nua&\simeq\biggl(\frac{(p-1)\pi^{\frac{3}{2}}3^{\frac{p+1}{2}}}{4}\frac{enR{\,\rm min}\big[\big(\frac{v}{v_{\rm DN}}\big)^2,\,1\big]}{\gamm^5B}\biggl)^{\frac{2}{p+4}}\num
	\label{eq nu_a}\\
&\simeq\begin{cases}
\big(4.0\times10^6{\,\rm Hz}\big)_{p=2.5}\,\bepen^{\frac{2}{p+4}}
\\
\,\,\,\,\,\,\,\,\varepsilon_{\rm B,-2}^{\frac{p+2}{2(p+4)}}n_0^{\frac{p+6}{2(p+4)}}v_9^{\frac{p+6}{p+4}}R_{17}^{\frac{2}{p+4}}&:\,v<v_{\rm DN},\\
\big(2.3\times10^6{\,\rm Hz}\big)_{p=2.5}\,\bepen^{\frac{2(p-1)}{p+4}}
\\
\,\,\,\,\,\,\,\,\varepsilon_{\rm B,-2}^{\frac{p+2}{2(p+4)}}n_0^{\frac{p+6}{2(p+4)}}v_{9}^{\frac{5p-2}{p+4}}R_{17}^{\frac{2}{p+4}}&:\,v_{\rm DN}<v.
\end{cases}
	\nonumber
\end{align}
Here in the second line we use $p=2.5$.
We emphasize that hereafter when we estimate the numerical values we adopt $p=2.5$ and write the results with parentheses like $()_{p=2.5}$.
Equations for general $p$ are given in Appendix \ref{append p}.
For $\nua>\num$, the synchrotron spectrum is given by
\begin{align}
F_{\nu}=\begin{cases}
F_{\num}(\nua/\num)^{\frac{1-p}{2}}(\num/\nua)^{5/2}(\nu/\num)^{2}&:\nu<\num, \\
F_{\num}(\nua/\num)^{\frac{1-p}{2}}(\nu/\nua)^{5/2}&:\num<\nu<\nua, \\
F_{\num}(\nu/\num)^{\frac{1-p}{2}}&:\nua<\nu. \\
\end{cases}
\end{align}
In particular the spectral peak is given by at $\nua$
\begin{align}
F_{\nua}&=F_{\num}(\nua/\num)^{\frac{1-p}{2}}
	\label{eq f nu_a}\\
&\simeq\begin{cases}
\big(5.4\times10^{-3}{\,\rm \mu Jy}\big)_{p=2.5}\,\bepen^{\frac{5}{p+4}}\varepsilon_{\rm B,-2}^{\frac{2p+3}{2(p+4)}}\\
\,\,\,\,\,\,\,\,n_0^{\frac{2p+13}{2(p+4)}}v_9^{\frac{2p+13}{p+4}}R_{17}^{\frac{2p+13}{p+4}}\big(\frac{\Omega}{4\pi}\big)d_{\rm L,27}^{-2}&:\,v<v_{\rm DN},\\
\big(1.3\times10^{-3}{\,\rm \mu Jy}\big)_{p=2.5}\,\bepen^{\frac{5(p-1)}{p+4}}\varepsilon_{\rm B,-2}^{\frac{2p+3}{2(p+4)}}\\
\,\,\,\,\,\,\,\,n_0^{\frac{2p+13}{2(p+4)}}v_{9}^{\frac{12p-7}{p+4}}R_{17}^{\frac{2p+13}{p+4}}\big(\frac{\Omega}{4\pi}\big)d_{\rm L,27}^{-2}&:\,v_{\rm DN}<v,
\end{cases}
	\nonumber
\end{align}
and the flux density in $\nua<\nu$ and $\num<\nu<\nua$, which are the relevant regimes in our study, are given by
\begin{align}
F_{\nu>\nua}&=F_{\nua}(\nu/\nua)^{\frac{1-p}{2}}
	\label{eq f nu1} \\
&\simeq\begin{cases}
\big(3.8\times10^{-5}{\,\rm \mu Jy}\big)_{p=2.5}\,\bepen\varepsilon_{\rm B,-2}^{\frac{p+1}{4}}\\
\,\,\,\,\,\,\,\,n_0^{\frac{p+5}{4}}v_9^{\frac{p+5}{2}}R_{17}^3\big(\frac{\Omega}{4\pi}\big)\nu_{\rm 3GHz}^{\frac{1-p}{2}}d_{\rm L,27}^{-2}&:\,v<v_{\rm DN},\\
\big(6.1\times10^{-6}{\,\rm \mu Jy}\big)_{p=2.5}\,\bepen^{p-1}\varepsilon_{\rm B,-2}^{\frac{p+1}{4}}\\
\,\,\,\,\,\,\,\,n_0^{\frac{p+5}{4}}v_{9}^{\frac{5p-3}{2}}R_{17}^3\big(\frac{\Omega}{4\pi}\big)\nu_{\rm 3GHz}^{\frac{1-p}{2}}d_{\rm L,27}^{-2}&:\,v_{\rm DN}<v,
\end{cases}
	\nonumber\\
F_{\nu<\nua}&=F_{\nua}(\nu/\nua)^{\frac{5}{2}}
	\label{eq f nu2} \\
&\simeq8.2\times10^4{\,\rm \mu Jy\,}\varepsilon_{\rm B,-2}^{-1/4}n_0^{-1/4}v_9^{-1/2}R_{17}^2\biggl(\frac{\Omega}{4\pi}\biggl)\nu_{\rm 3GHz}^{5/2}d_{\rm L,27}^{-2},
	\nonumber
\end{align}
respectively, where $\nu_{\rm 3GHz}=\nu/3\,\rm{GHz}$.
Note that the flux density for $\num<\nu<\nua$ (Eq. \ref{eq f nu2}) has a common dependence on the parameters for the both regimes.

\subsection{Radio constraints}
\begin{figure}
\begin{center}
\includegraphics[width=85mm, angle=0]{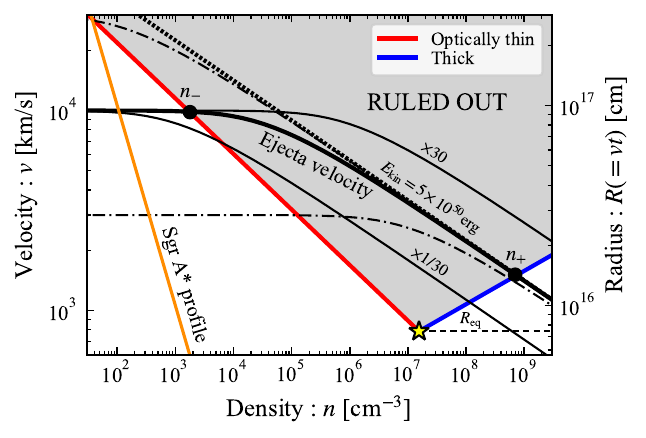}
\caption{Example of the constraint on the density and velocity space by a typical radio upper-limit of $F_{\nu}=30\,\rm \mu Jy$ at $\nu=3\,\rm GHz$ for $z=0.072$ ($d_{\rm L}=10^{27}\,\rm cm$) and $t=3\,\rm yr$.
Red and blue lines represent the boundaries of the excluded (gray) region imposed by the optically thin ($\nu>\nua$) and thick ($\nu<\nua$) regimes, respectively.
The adopted parameters are $\bepe=0.1$, $\epb=0.01$, $p=2.5$, and $\Omega=4\pi$.
The intersection of two lines (star) gives the minimal velocity $v_{\rm eq}$ corresponding to the equipartition radius $R_{\rm eq}(=v_{\rm eq}t)$.
The thick-black-solid curve denotes the trajectory of shock velocity given by solving Eq. \eqref{eq conservation} for $M_{\rm ej}=0.5\,\Msun$ and $v_{\rm in}=10^{4}\,\rm km\,s^{-1}$.
The thin-black-solid curves show trajectories with the same $v_{\rm in}$ but 30 times larger and smaller $E_{\rm kin}$ (or equivalently $M_{\rm ej}$).
The black dash-dotted curves represent trajectories with the same $E_{\rm kin}$ but 3 times higher and lower $v_{\rm in}$.
For large density, these curves with the same $E_{\rm kin}$ approach to the black-dotted line $v\propto n^{-1/5}$, which corresponds to decelerating jets.
Within the outflow model, any density profile intersecting the trajectory within the ruled-out region for $n_{-}\leq n \leq n_+$ is excluded.
The density profile of Sgr A* (orange line, $n\simeq10\,{\rm cm^{-3}}(R/10^{18}\,\rm cm)^{-1}$, \citealt{Baganoff+2003,Gillessen+2019}) is allowed by the model.
}
\label{fig limit sample}
\end{center}
\end{figure}

We constrain the outflow's properties such as velocity $v$ and solid angle $\Omega$ and the CNM density $n$ by using the radio observations.
The radius of the outflow is approximately estimated by $R\simeq vt$.
Strictly speaking, the radius is given not by the outflow velocity $v$ but by the shock velocity, which is slightly larger than $v$.
This approximation holds even after the outflow starts to decelerate.
For instance, with a density profile of $n\propto R^{-k}$ ($k<3$), the radius and velocity evolve as $R\propto t^{2/(5-k)}$ and $v\simeq dR/dt$, respectively, which gives $R\simeq [(5-k)/2]vt$. These numerical factors do not change our results significantly as long as we take $v$ as a fundamental quantity (instead of $R$).
Here $t$ is the time measured since the outflow launch.
Therefore, the synchrotron flux is determined by three key parameters of $v$, $\Omega$, and $n$.
By substituting $R=vt$ to Eqs. \eqref{eq f nu1} and \eqref{eq f nu2} and setting the flux smaller than the upper limit $F_\nu$ at frequency $\nu$, we constrain the combinations of the parameters for optically thin and the deep-Newtonian phase ($v<v_{\rm DN}$),
\begin{align}
n_{0}^{\frac{p+5}{4}}v_9^{\frac{p+11}{2}}\Omega\lesssim\big(3.2\times10^{8}\big)_{p=2.5}\,\bepen^{-1}\varepsilon_{\rm B,-2}^{-\frac{p+1}{4}}t_{\rm yr}^{-3}\nu_{\rm 3GHz}^{\frac{p-1}{2}}d_{\rm L,27}^{2}F_{\rm 30\mu Jy},
	\label{eq limit f dn}
\end{align}
for optically thin and $v>v_{\rm DN}$ case,
\begin{align}
&n_0^{\frac{p+5}{4}}v_{9}^{\frac{5p+3}{2}}\Omega \lesssim \big(2.0\times10^{9}\big)_{p=2.5}\,\bepen^{1-p}\varepsilon_{\rm B,-2}^{-\frac{p+1}{4}}t_{\rm yr}^{-3}\nu_{\rm 3GHz}^{\frac{p-1}{2}}d_{\rm L,27}^{2}F_{\rm 30\mu Jy},
	\label{eq limit f}
\end{align}
and for the optically thick case,
\begin{align}
&n_0^{-1/4}v_9^{3/2}\Omega \lesssim4.6\times10^{-3}\,\varepsilon_{\rm B,-2}^{1/4}t_{\rm yr}^{-2}\nu_{\rm 3GHz}^{-5/2}d_{\rm L,27}^{2}F_{\rm 30\mu Jy},
	\label{eq limit g}
\end{align}
where $t_{\rm yr}=t/\rm yr$ and $F_{30\mu\rm Jy}=F_\nu/\rm 30\,\mu Jy$.
In particular, the velocity is constrained by
\begin{align}
v\lesssim\begin{cases}
\big(1.8\times10^5{\,\rm km\,s^{-1}\,}\big)_{p=2.5}\,\Big[\bepen^{-1}\varepsilon_{\rm B,-1}^{-\frac{p+1}{4}}\\
\,\,\,t_{\rm yr}^{-3}\nu_{\rm 3GHz}^{\frac{p-1}{2}}d_{\rm L,27}^2F_{\rm 30\mu Jy}n_0^{-\frac{p+5}{4}}\Omega^{-1}\Big]^{\frac{2}{p+11}}&: {\rm thin\,}(v<v_{\rm DN}),\\
\big(1.6\times10^{5}{\,\rm km\,s^{-1}\,}\big)_{p=2.5}\,\Big[\bepen^{1-p}\varepsilon_{\rm B,-2}^{-\frac{p+1}{4}}\\
\,\,\,t_{\rm yr}^{-3}\nu_{\rm 3GHz}^{\frac{p-1}{2}}d_{\rm L,27}^{2}F_{\rm 30\mu Jy}n_0^{-\frac{p+5}{4}}\Omega^{-1}\Big]^{\frac{2}{5p+3}}&: {\rm thin\,}(v>v_{\rm DN}),\\
1.3\times10^{3}{\,\rm km\,s^{-1}\,}\,\varepsilon_{\rm B,-2}^{1/6}t_{\rm yr}^{-4/3}\nu_{\rm 3GHz}^{-5/3}\\
\,\,\,d_{\rm L,27}^{4/3}F_{\rm 30\mu Jy}^{2/3}n_0^{1/6}\Omega^{2/3}&: {\rm thick},
\end{cases}
    \label{eq velocity limit}
\end{align}
corresponding to Eqs. \eqref{eq limit f dn}, \eqref{eq limit f}, and \eqref{eq limit g}, respectively.
Fig. \ref{fig limit sample} depicts an example demonstrating how a radio upper-limit constrains the density and velocity space for a given $\Omega$.

The spectrum peaks at $\nua$ and this implies that regardless of the external density, a minimal velocity is required to realize a given flux for a given $\Omega$.
The minimal velocity is obtained by equating the observed flux (upper limit) and frequency with $F_{\nua}$ and $\nua$ (Eqs. \ref{eq nu_a} and  \ref{eq f nu_a}), or equivalently given by the intersection of the velocity limits at the optically thin and thick regimes (Eqs. \ref{eq velocity limit}).
The minimal velocity and the corresponding density are given by
\begin{align}
v_{\rm eq}&\simeq\begin{cases}
\big(8.3\times10^3{\,\rm km\,s^{-1}\,}\big)_{p=2.5}\,\,\bepen^{-\frac{1}{2p+13}}\varepsilon_{\rm B,-2}^{\frac{1}{2p+13}}\\
\,\,\,\,\,\,t_{\rm yr}^{-1}\nu_{\rm 3GHz}^{-1}d_{\rm L,27}^{\frac{2(p+6)}{2p+13}}F_{\rm 30\mu Jy}^{\frac{p+6}{2p+13}}\Omega^{-\frac{p+6}{2p+13}}&:v<v_{\rm DN},\\
\big(9.2\times10^3{\,\rm km\,s^{-1}\,}\big)_{p=2.5}\,\bepen^{\frac{1-p}{4p+9}}\varepsilon_{\rm B,-2}^{\frac{1}{4p+9}}\\
\,\,\,\,\,\,t_{\rm yr}^{-\frac{2p+13}{4p+9}}\nu_{\rm 3GHz}^{-\frac{2p+13}{4p+9}}d_{\rm L,27}^{\frac{2(p+6)}{4p+9}}F_{\rm 30\mu Jy}^{\frac{p+6}{4p+9}}\Omega^{-\frac{p+6}{4p+9}}&:v>v_{\rm DN}.
\end{cases}
    \label{eq v eq}\\
n_{\rm eq}&\simeq\begin{cases}
\big(6.8\times10^4{\,\rm cm^{-3}}\big)_{p=2.5}\bepen^{-\frac{6}{2p+13}}\varepsilon_{\rm B,-2}^{-\frac{2p+7}{2p+13}}\\
\,\,\,\,\,\, t_{\rm yr}^2\nu_{\rm 3GHz}^4d_{\rm L,27}^{-\frac{4(p+8)}{2p+13}}F_{\rm 30\mu Jy}^{-\frac{2(p+8)}{2p+13}}\Omega^{\frac{2(p+8)}{2p+13}}&:v<v_{\rm DN},\\
\big(1.3\times10^5{\,\rm cm^{-3}\,}\big)_{p=2.5}\bepen^{\frac{6(1-p)}{4p+9}}\varepsilon_{\rm B,-2}^{-\frac{4p+3}{4p+9}}\\
\,\,\,\,\,\, t_{\rm yr}^{\frac{2(10p-3)}{4p+9}}\nu_{\rm 3GHz}^{\frac{4(7p+3)}{4p+9}}d_{\rm L,27}^{-\frac{20p}{4p+9}}F_{\rm 30\mu Jy}^{-\frac{10p}{4p+9}}\Omega^{\frac{10p}{4p+9}}&:v>v_{\rm DN},
\end{cases}
	\label{eq n eq}
\end{align}
respectively.
At $R_{\rm eq}\equiv v_{\rm eq}t$ and $n_{\rm eq}$ the upper limit corresponds to the spectral peak at $\nu_{\rm a}$.
Hence in particular for the case of $v_{\rm eq}<v_{\rm DN}$ the radius $R_{\rm eq}$ is comparable to the equipartition radius \citep[e.g.][]{Chevalier1998,BarniolDuran+2013}.
Strictly speaking, 
our ``equipartition'' radius  minimizes the total energy in the emitting site (including all electrons' energy).
Note we allow for a general relations $\varepsilon_{\rm B}\ll\varepsilon_{\rm e}$ \citep[see also][]{Chevalier1998} while the literal equipartition is realized with $\varepsilon_{\rm B}=(6/11)\varepsilon_{\rm e}$ at the equiaprtition radius of \cite{BarniolDuran+2013}.
This of course results in that our total energy is slightly larger, but the radius remains practically unchanged.

There is more important difference between our method and the usual application of the equipartition method \citep{Chevalier1998,BarniolDuran+2013}.
The equipartition method is applicable only when the spectral peak is observed.
On the other hand, our formulation can be applied to any observations.

Additionally the density $n_{\rm eq}$ that we find is typically 30-100 times larger than the one obtained by the equipartition method.
This is because we take into account the fact that only a fraction $(v/v_{\rm DN})^2\simeq1-3\,\%$ of the electrons participates in the power-law distribution in the deep-Newtonian phase, that is usually not considered in the simple version of the equipartition calculations.

\subsection{Outflow model}\label{sec outflow model}
Consider an outflow launched into a solid angle $\Omega$.
Generally, the outflow is ejected with a range of velocities and we denote the mass moving at larger velocity than $v$ as $M_{\rm ej}(>v)$ and its corresponding kinetic energy as $E_{\rm kin}(>v)$.
At a given moment, the shock velocity of the outflow $v$ is determined by the energy conservation \citep{Piran+2013}:
\begin{align}
\Ekin(>v)&=\big[\Mej(>v)+ M(R)\big]v^2/2,
	\label{eq conservation}
\end{align}
where the swept-up CNM mass is approximated as in Eq. \eqref{eq electron number}:
\begin{align}
M(R)\simeq\Omega \MP n R^3.
	\label{eq ism mass}
\end{align}
The shock radius is reasonably given by $R\simeq vt$.

As a simple example, we consider an outflow characterized by a single initial velocity $v_{\rm in}$ with mass $M_{\rm ej}$, kinetic energy $E_{\rm kin}=M_{\rm ej}v_{\rm in}^2/2$, and angle $\Omega=4\pi$.
Fig. \ref{fig limit sample} depicts trajectories of velocity obtained by solving Eq. \eqref{eq conservation} for different densities at the shock front.
For small CNM density, the outflow is still traveling with its initial velocity (free expansion).
When the swept-up CNM mass is larger than the outflow's mass, its velocity decreases.
The trajectory asymptotes to a line, $v\propto n^{-1/5}$ which is obtained by neglecting the outflow mass ($\Mej\to0$). As we will see in \S \ref{sec jet}, this line represents the trajectory of decelerating jets and its normalization is determined only by the outflow's kinetic energy.

When the outflow velocity is larger than the minimal velocity $v_{\rm eq}$ corresponding to a given upper limit on $F_{\nu}$, the trajectory intersects the excluded region for $n_{-}<n<n_{+}$.
Here we define $n_+$ and $n_-$ as the densities at the intersections with the optically thick and thin boundaries, respectively (see Fig. \ref{fig limit sample}).
We find that for most cases, the density $n_{+}$ is much larger than the relevant CNM-density range.
Hence we only consider the branch of $n<n_{-}$, where $n_{-}$ gives an upper limit on the density.
As long as we consider the density limit in the optically thin regime $n_{-}$, the outflow velocity does not change significantly from the initial value $v_{\rm in}$.
Therefore the limiting density depends on $\varepsilon_{\rm B}$, $v_{\rm in}$ and $\Omega$ as
\begin{align}
n_{-}\propto \varepsilon_{\rm B}^{-\frac{p+1}{p+5}}v_{\rm in}^{-\frac{2(p+11)}{p+5}}\Omega^{-\frac{4}{p+5}}\simeq \varepsilon_{\rm B}^{-0.47}v_{\rm in}^{-3.6}\Omega^{-0.53},
    \label{eq density limit para}
\end{align}
where the last equality holds for $p=2.5$.
 
In the following calculations we adopt $\bar{\varepsilon}_{\rm e}=0.1$, $\varepsilon_{\rm B}=0.01$, $p=2.5$ as the fiducial values (several radio detected events have different $p$, see Table \ref{table detection}).
The value of $\varepsilon_{\rm B}$ is not well constrained by the observations and can vary for different events shifting the boundary of the ruled-out region in Fig. \ref{fig limit sample}.

To summarize this section we note that when a radio signal is detected this immediately gives a lower limit on the velocity $v_{\rm eq}$ or equivalently the combination of $v \Omega^{a}$ where $a=(p+6)/(2p+13)\simeq0.47$ for $p=2.5$ ($v<v_{\rm DN}$) or $a=(p+6)/(4p+9)\simeq 0.45$ for $p=2.5$ ($v>v_{\rm DN}$).
This limit is independent of the density.
For an upper limit on the radio flux, we obtain a forbidden region in $n$ and $v$ space as a function of $\Omega$.
Given an outflow model, we can constrain the density.
Alternatively, once a density profile is specified, we can constrain the outflow's properties such as $v$ and $\Omega$.

A give outflow is characterized by its geometry, mass, and velocity (actually by $M_{\rm ej}(>v)$).
Among these three parameters, with a detection we have a direct bound on the product of two $v \Omega^a$ (as described above). The total mass is less important as typically only a small fraction of it is sufficient to power the observed signal. However, as we show later its velocity distribution might be critical.

Going back to Fig. \ref{fig limit sample}, we note that the density and velocity in the shaded region are ruled out for an upper limit.
For a detection without an SSA spectral peak, its parameters should be on the boundary of the ruled-out region.
Only when the spectral peak is detected the parameters are determined on the bottom of the boundary (a star), which is the usual equipartition method.

\section{Application to Astrophysical Models}\label{sec app}
We turn now to apply the method developed in \S \ref{sec method} to four different possible sources of the flare.
These are distinguished by their geometry, typical masses, and expected velocities.

Disk winds, that arise if circularization takes place rapidly and a super-Eddington emitting disk forms, have a quasi spherical geometry.
The unbound disrupted stellar mass has a wedge-like geometry while a possible Newtonian jet will have a conical shape.
Among the three only for the disrupted stellar mass we have estimates of the mass-velocity distribution, $M_{\rm ej}(>v)$.
For the disk wind we can expect the velocity to be of order of (or larger than) the escape velocity from the photosphere of several thousand $\rm km\,s^{-1}$ \citep[e.g.][]{Matsumoto&Piran2021}.
For Newtonian jets considered within this context of TDE radio emission \citep[see e.g.][]{Alexander+2016}, its mass, velocity, and even the opening angle are typically not constrained by other considerations.  

\subsection{Spherical outflow - Disk Wind} \label{sec wind}
First we consider a spherical outflow.
Such an outflow can arise from a disk wind.
If circularization is efficient, a compact accretion disk forms rapidly after the bound debris returns to the pericenter.
Since the mass fallback rate is well above the Eddington rate, a strong disk wind could emerge and carry out a significant mass \citep{Blandford&Begelman1999}.
The wind plays a role of an envelope surrounding the disk and absorbing the disk's X-rays to reprocess to optical light \citep{Loeb&Ulmer1997,Metzger&Stone2016,Roth+2016}.
Recently we have shown that the observations imply that such winds are too massive and would carry out more than the available mass (\citealt{Matsumoto&Piran2021}, see also \citealt{Uno&Maeda2020b}).
Naturally a radio signal is expected from the interaction of such a wind with the CNM and null detection of radio emission from TDEs provides an independent constrain on this scenario.

Consider an isotropic ($\Omega=4\pi$) outflow characterized by a single (initial) velocity $v_{\rm in}=10^4\,\rm km\,s^{-1}$ and mass $\Mej=0.5\,\Msun$ (as we discuss later as the energy emitted by the shock is very small, the outcome does not depend on the total mass as long as it is sufficiently larger than the swept-up mass, $M_{\rm ej}>M(R)$).
These values follow from the reprocessed model \citep[e.g.][]{Metzger&Stone2016} and considered in \S \ref{sec outflow model} as an example.
The shock radius is approximately given by $R\simeq vt$, and the time is measured since the wind launch.
For $t$, we use, here and in the rest of the paper,  the time measured since the discovery of TDEs, which is a reasonable approximation because the radio observation times are typically late.
For a given time and CNM density at the shock front, the shock velocity is given by Eq. \eqref{eq conservation}.
Fig. \ref{fig limit sample} depicts trajectories of the outflow velocity with the corresponding excluded parameter space obtained by a typical radio upper-limit of $F_{\nu}=30\,\mu \rm Jy$ at $\nu=3\,\rm GHz$ for $z=0.072$ ($d_{\rm L}=10^{27}\,\rm cm$) and $t=3\,\rm yr$.

\begin{figure}
\begin{center}
\includegraphics[width=85mm, angle=0]{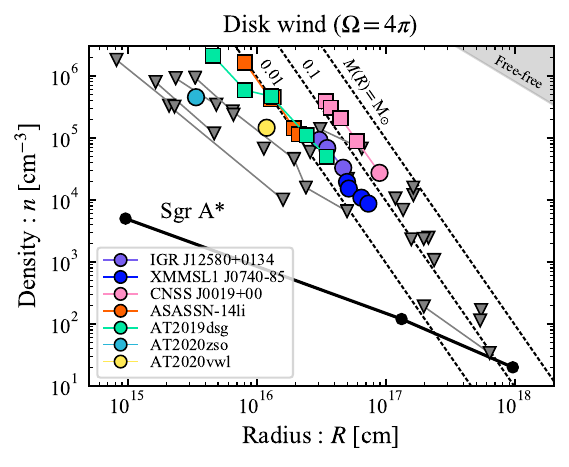}
\caption{
Limits or required CNM density by the radio observations for the case of a spherical outflow ($\Omega=4\pi$, $M_{\rm ej}=0.5\,\Msun$, and $v_{\rm in}=10^4\,\rm km\,s^{-1}$).
Down-triangles, circles, and squares represent the upper limits on the density by radio limits, required density by radio TDEs without spectral peak, and density obtained by the equipartition method, respectively.
Upper limits at large radius correspond to the radio limits at late time. 
The dashed lines show the locations where the enclosed masses are $M(R)=10^{-3}-10^{-1}\,\Msun$.
If the outflow mass is larger than $M(R)$, it does not decelerate significantly.
The black line represents the density profile of Sgr A* (\citealt{Baganoff+2003,Gillessen+2019}).
In the upper right shaded region, the free-free absorption reduces the observed flux (assumed the virial temperature for $M_{\rm BH}=10^{6.5}\,\Msun$) at $\nu=1\,\rm GHz$.
}
\label{fig limit win2}
\end{center}
\end{figure}

For each radio upper-limit in Fig. \ref{fig lc}, we calculate the quantities $v_{\rm eq}$, $n_{\rm eq}$, $v_{-}$, and $n_{-}$ (corresponding to the optically thin regime) assuming $p=2.5$ and tabulate them in Table \ref{table limit}.
Fig. \ref{fig limit win2} depicts the upper limits on the density $n_{-}$ at the distance $R=v_{-}t$ when the limiting velocity is larger than the critical velocity $v_{-}>v_{\rm eq}$.
Due to the large solid angle, which minimizes the minimal-required velocity $v_{\rm eq}$, we obtain meaningful and strongest constraints (see below for the cases of $\Omega<4\pi$). 
It is the late upper-limits ($>10\,\rm yr$) that give the strongest limits (furthest away from the BH and hence the expected density is lowest).
As seen in Eq. \eqref{eq density limit para}, increasing the velocity by a factor of 2, which is the case of the high tail of the velocity distribution, will decrease the density limit about ten times. 
Limits are meaningful compared to the density profile seen in the radio-detected TDEs.
These results do not rule out the strong disk wind scenario such as the reprocessed optical-emission model.

For radio-detected TDEs, the wind interaction with the CNM can be an efficient radio source provided that its velocity is not too small ($v_{\rm in}\gtrsim10^{4}\,\rm km\,s^{-1}$).
For three radio TDEs with a spectral peak (CNSS J0019+00, ASASSN-14li, and AT2019dsg), we estimate the densities $n_{\rm eq}$ corresponding to the minimal velocities $v_{\rm eq}$.
For these events, observations at different times show that the outflow does not decelerate  (see Table \ref{table detection}).
This, in turn, suggests the outflow mass is larger than the swept-up CNM mass: $M_{\rm ej}\gtrsim M(R)\simeq0.01-0.1\,\Msun$ (as shown in Fig. \ref{fig limit win2}).
Hence the outflow's kinetic energy should be larger than $E_{\rm kin}\gtrsim v_{\rm eq}^2M(R)/2\simeq10^{49-50}\,\rm erg$.

The densities that we find here are 30-100 times larger than those obtained by previous works based on the equipartition method \citep{Alexander+2016,Krolik+2016,Anderson+2020,Stein+2021,Cendes+2021b}, because these works considered only the number of relativistic electrons, which is typically smaller by a factor of $(v/v_{\rm DN})^2$ than the total electron density in the deep-Newtonian phase.\footnote{For ASASSN-14li, our density is further 10 times larger than that given by \cite{Alexander+2016}.}
Accordingly their estimates of outflow's mass and kinetic energy are smaller than ours by the similar factor.
The number density obtained by \cite{Yalinewich+2019b} for ASASSN-14li (after correcting the difference of solid angle) is also slightly ($\sim3$ times) smaller than ours because they assumed the fraction of accelerated electrons $10\,\%$.

We carry out similar calculations for radio-detected TDEs without a spectral peak. 
Assuming an initial velocity $v_{\rm in}=10^4\,\rm km\,s^{-1}$ we find that the wind velocity does not vary during the observations.
The outflow mass is hence bounded by the swept-up mass and can be small such as $M_{\rm ej}\gtrsim M(R)\simeq0.03\,\Msun$ and the minimal energy becomes $E_{\rm kin}\gtrsim3\times10^{49}{\,\rm erg\,}M_{\rm ej,-1.5}v_{\rm in,9}^2$. 
Some of the radio-detected TDEs require densities larger  than those of  radio upper-limits TDEs.
This suggests that either not every TDE is accompanied by a disk wind or the CNM profiles vary significantly among galaxies.

\subsection{Wedge geometry - Unbound debris}
\label{sec unbound}
\subsubsection{Dynamics of unbound debris}
About half of the stellar mass torn apart in a TDE is ejected as unbound debris.
The interaction of the debris with the CNM should produce radio emission.
We turn, now, to constrain the debris properties and the CNM density using the radio observations.

\begin{figure}
\begin{center}
\includegraphics[width=85mm, angle=0]{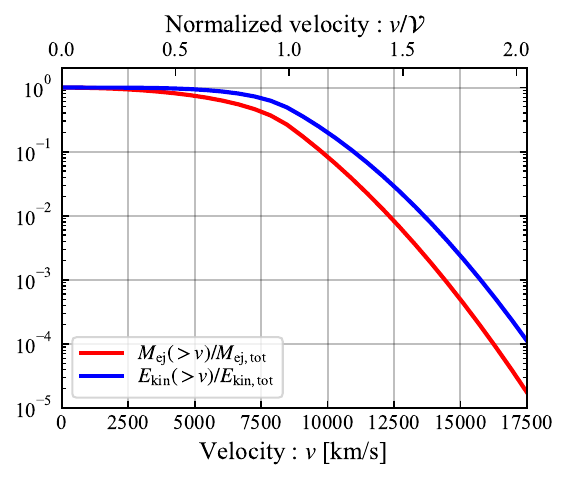}
\caption{Cumulative mass and kinetic energy of unbound debris (reconstructed from the specific energy distribution, Eq. \ref{eq un profile}) for parameters of $M_{\rm *}=\Msun$ ($R_*=R_\odot$), $M_{\rm BH}=10^{6.5}\,\Msun$, and $\alpha=3$. The velocity normalization $\cal V$ is given by Eq. \eqref{eq un velocity}.}
\label{fig profile un}
\end{center}
\end{figure}

We consider a disruption event of a star with mass $M_*$ and radius $R_*$ by a BH with mass $M_{\rm BH}$.
After the disruption, the stellar debris has a flat distribution over specific energy within a characteristic energy of $\Delta \epsilon\equiv (GM_{\rm BH} R_*/R_{\rm T}^2)\Xi$, where $G$ is the gravitational constant and $R_{\rm T}\equiv R_*(M_{\rm BH}/M_*)^{1/3}$ is the tidal radius.
$\Xi$ is a numerical factor derived by \cite{Ryu+2020} in order to include corrections arising due to the internal stellar  structure and relativistic effects.
This correction is less than a factor of $2$ for typical values and it becomes $\Xi\simeq1.3$ for a star with $M_{\odot}$ and BH with $M_{\rm BH}=10^{6.5}\,\Msun$.
The corresponding typical velocity is
\begin{align}
{\cal V}=\sqrt{2\Delta\varepsilon}\simeq8600{\,\rm km\,s^{-1}\,}R_{*,0}^{-1/2}M_{*,0}^{1/3}M_{\rm BH,6.5}^{1/6}(\Xi/1.3)^{1/2}\ ,
	\label{eq un velocity}
\end{align}
where we normalize the radius and mass by solar values.
As a zeroth order approximation we could consider outflow with mass $M_{\rm ej}\simeq0.5\,\Msun$ and an opening angle of $\Omega\sim0.1$ (as discussed later) with this velocity.
However, \cite{Ryu+2020} also found the detailed specific-energy distribution.
This distribution has a tail beyond $\Delta \epsilon$ whose shape can be approximated by an exponential.
Since the small fraction of fast unbound debris can contribute to or even dominate a radio flare \citep{Krolik+2016,Yalinewich+2019b}, we take this structure into account and adopt the following distribution
\begin{align}
\frac{dM}{d\epsilon}&=\frac{\alpha M_*}{2(\alpha+1)\Delta\epsilon}\begin{cases}
1&:\epsilon<\Delta\epsilon \ , \\
\exp\Big[-\alpha\Big(\frac{\epsilon-\Delta\epsilon}{\Delta\epsilon}\Big)\Big]&:\epsilon>\Delta\epsilon \ , \\
\end{cases}
	\label{eq un profile}
\end{align}
where the slope of the exponential tail $\alpha\gtrsim3$ depends on the type of the disrupted star \citep{Ryu+2020} and we use $\alpha=3$ as a fiducial value.
The normalization is determined so that the total unbound mass for $\epsilon>0$ becomes $M_*/2$.
With a relation $\epsilon=v^2/2$, the debris mass and kinetic energy distributions over velocity are given by $d\Mej/dv=v(dM/d\epsilon)$ and $d\Ekin/dv=(v^2/2)(d\Mej/dv)$, respectively.
Fig. \ref{fig profile un} depicts the cumulative mass and kinetic energy profiles of the debris, $M_{\rm ej}(>v)$ and $E_{\rm kin}(>v)$.
The total kinetic energy is given by $\Ekin\simeq2.6\times10^{50}{\,\rm erg\,}R_{*,0}M_{*,0}^{5/3}M_{\rm BH,6.5}^{1/3}$ for $\alpha=3$.
Only a small fraction of debris has a larger velocity $v\gtrsim10^4\,\rm km\,s^{-1}$.

The dynamics of the unbound debris is determined by using Eq. \eqref{eq conservation} and the mass and energy distributions.
The debris expands in a wedge geometry like a fan \citep[constant ratio of the width to radius $H/R$,][]{Strubbe&Quataert2009,Yalinewich+2019b} producing a bow shock with a typical solid angle of $\Omega\sim0.1$ \citep{Yalinewich+2019b}.
At first, the faster and less massive debris travels ahead of the whole debris and its bow shock dominates the radio emission.
As it decelerates, more massive debris dominates.
At a given moment, the emission mainly comes from the bow shock formed by the debris which just begins to decelerate.
The shock velocity is comparable to the debris velocity.

\begin{figure}
\begin{center}
\includegraphics[width=85mm, angle=0]{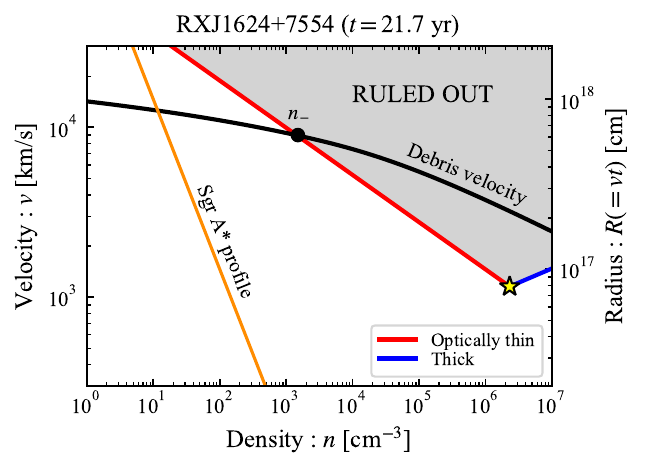}
\caption{The same as Fig. \ref{fig limit sample} but for the radio upper-limit of RXJ1624+7554 at $t=21.6\,\rm yr$ in the case of unbound debris with a solid angle $\Omega=0.1$.
The black curve denotes the velocities of the unbound debris and its bow shock as a function of the density given by Eq. \eqref{eq conservation} with the profile of Eq. \eqref{eq un profile}, for given debris mass $M_{\rm ej}=0.5\,\Msun$, kinetic energy $E_{\rm kin}\simeq3\times10^{50}\,\rm erg$, and BH mass $M_{\rm BH}=10^{6.5}\,\Msun$.
}
\label{fig limit un}
\end{center}
\end{figure}

Fig. \ref{fig limit un} depicts the debris velocity as a function of the density at the shock front.
The energy and mass profiles are calculated by the distribution given by Eq. \eqref{eq un profile} (see also Fig. \ref{fig profile un}).
We approximate the radius $R\simeq vt$.
As the debris sweeps up CNM and decelerates, it expands sideways and increases the solid angle of the bow shock $\Omega$.
The expanding part sweeps up more CNM and decelerates faster.
We do not include this sideways expansion in our estimate  and we assume the constant angle because the rapidly decelerating part does not contribute to the emission significantly (note that in the optically thin regime the flux sensitively depends on the velocity, Eq. \ref{eq f nu1} and see an argument in \citealt{Margalit&Piran2015}).

\subsubsection{Observational constraints}
We derive the constraint in the context of unbound debris.
Fig. \ref{fig limit un} shows the ruled-out parameter space for the null detection in RXJ1624+7554 at $t=21.6\,\rm yr$ (shown in the first line of Table \ref{table limit}).
In this example, any velocity below  $v_{\rm eq}\simeq10^3\,\rm km\,s^{-1}$ is allowed (for this assumed $\Omega$). 
The velocity trajectory overlaps the ruled-out region and the flux upper-limit requires the density smaller than $n<n_-\simeq10^3\,\rm cm^{-3}$.

\begin{figure}
\begin{center}
\includegraphics[width=85mm, angle=0]{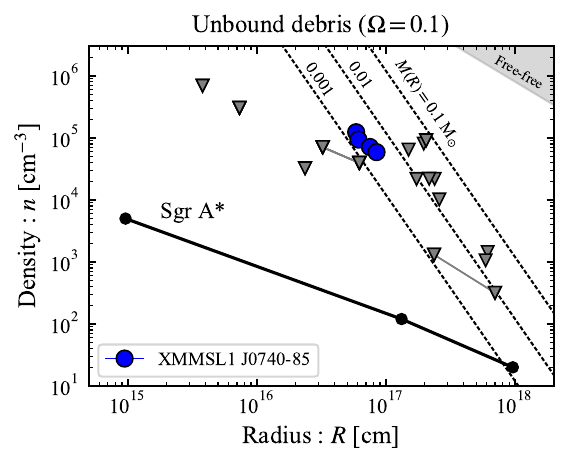}
\caption{The same as Fig. \ref{fig limit win2} but for the case of unbound debris ($\Omega=0.1$, $M_*=\Msun$, and $M_{\rm BH}=10^{6.5}\,\Msun$).
The dashed lines show the locations where the enclosed masses are $M(R)=10^{-3}-10^{-1}\,\Msun$ (see Eq. \ref{eq ism mass}).}
\label{fig limit un2}
\end{center}
\end{figure}

Fig. \ref{fig limit un2} depicts the resulting upper limits on the density $n_{-}$.
In order to calculate the debris velocity, we adopt the same parameters of $M_*=\Msun$, $R_*=R_{\odot}$, and $M_{\rm BH}=10^{6.5}\,\Msun$.
Using the same BH mass for all events is reasonably justified because the characteristic velocity weakly depends on this parameter (see Eq. \ref{eq un velocity}).
With this fiducial debris model, only half of the TDEs with upper limits gives meaningful constraints on the density. 
These density limits are all larger than the CNM density in Sgr A*.
The other half does not produce a radio flare as bright as the upper limit because its velocity is smaller than the minimal one $v_{\rm eq}$.

For TDEs with a spectral peak (CNSS J0019+00, ASASSN-14li and AT2019dsg), if the peak is caused by SSA, the velocity and density should be equal to $v_{\rm eq}$ and $n_{\rm eq}$ as obtained by the equipartition method.
However, with the solid angle of $\Omega=0.1$, the minimal velocities are typically larger than that of the unbound debris $v_{\rm eq}\gtrsim10^4\,\rm km\,s^{-1}$. 
With these parameters of our fiducial debris model the unbound debris cannot be the radio source in these events.  
Similarly, for radio TDEs without a spectral peak only XMMSL J0740-85 can be powered by unbound debris in our model
 (see Fig. \ref{fig limit un2}).
These limits are slightly relaxed if we take into account that the shock velocity is larger than the fluid velocity. But this is not sufficient to fully resolve the discrepancy.

Generally speaking, the radio emission from unbound debris is much dimmer than the other outflow components because of its small solid angle, $\Omega\sim0.1$.
The angle can be ten times larger than the fiducial value, $\Omega\sim1$ for TDEs in which the stellar pericenter is smaller than the tidal radius as studied by \cite{Yalinewich+2019b} for modeling of ASASSN-14li.
In such TDEs, the luminosity becomes $\sim10$ times larger and the equipartition velocity is also reduced by a factor three, $v_{\rm eq}\propto \Omega^{-\frac{p+6}{2p+13}}\simeq\Omega^{-0.47}$ for $p=2.5$, which results in more events give meaningful constraints on the density.
With this scaling, we find that debris should have a distribution in which at least $\sim10^{-2}\,\Msun$ of mass moves at $\gtrsim2\times10^{4}\,\rm km\,s^{-1}$.

\subsection{Conical outflow - a Newtonian Jet}\label{sec conical}
We turn now to conical outflows that are launched with sub-relativistic velocities \citep[e.g.][]{Alexander+2016}.
Such a jet-like structure can arise, for example, from a disk wind that has a strong angular structure.
Lacking a specific model for generation of such a jet, here, unlike the previous two cases we do not have specific model parameters to compare with and all that can be done is to infer jet parameters that fit the data.
Naturally, this is almost always possible. 
Relativistic jets behave differently and we consider those in the next section.

As the jet propagates in the surrounding CNM its energy is dissipated at the head and a hot cocoon forms \citep[see e.g.][]{Begelman&Cioffi1989,Bromberg+2012}.
The effective width of the emitting region of the jet is therefore  larger, but comparable, to its original width. Namely,  $\Omega\gtrsim2\pi\theta^2$, where $\theta$ is the original jet half-opening angle. 

As long as jet's mass is larger than the swept-up CNM mass, it does not decelerate. In this case we can simply follow the analysis in \S \ref{sec wind} using the assumed opening angle $\Omega$ instead of $4 \pi$ and obtain constraints on the velocity or density.
However,  we cannot simply scale the relations from the spherical analysis because in that case
we are usually in the deep-Newtonian regime. 
Due to the smaller $\Omega$, the required velocities in jets are much higher than those required for a spherical outflow and typically they are not in this deep-Newtonian regime.
We denote the minimal velocity, density constraint, and corresponding velocity for the spherical outflow ($\Omega=4\pi$ and in the deep-Newtonian phase, see Table \ref{table limit}) with a superscript $4\pi$ ($v_{\rm eq}^{4\pi}$, $n_-^{4\pi}$, and $v_-^{4\pi}$ respectively).
As seen in Eq. \eqref{eq v eq}, $v_{\rm eq}$ scales as 
\begin{align}
v_{\rm eq}&=v_{\rm DN}\biggl(\frac{v_{\rm DN}}{v_{\rm eq}^{4\pi}}\biggl)^{-\frac{2p+13}{4p+9}}\biggl(\frac{\Omega}{4\pi}\biggl)^{-\frac{p+6}{4p+9}}.
    \label{eq scale v eq}
\end{align}
The density limit scales with $\Omega$ and $v_{\rm in}(>v_{\rm DN})$ as
\begin{align}
n_{-}=n_{-}^{4\pi}\biggl(\frac{v_{\rm DN}}{v_{-}^{4\pi}}\biggl)^{-\frac{2(p+11)}{p+5}}\biggl(\frac{v_{\rm in}}{v_{\rm DN}}\biggl)^{-\frac{2(5p+3)}{p+5}}\biggl(\frac{\Omega}{4\pi}\biggl)^{-\frac{4}{p+5}}. 
    \label{eq scale density}
\end{align}
As the limits for  jets arise from  these scaling laws of the spherical limits we do not list them in a different column in Table \ref{table limit}.

\cite{Alexander+2016} considered a Newtonian jetted outflow as a radio source of ASASSN-14li. 
Adopting $\Omega\simeq0.3$ ($\theta\simeq0.33\,\rm rad\simeq13^\circ$) they find that the jet velocity should be $v_{\rm eq}\sim 0.1\,c$ (six times larger than the limit on a spherical freely-expanding outflow), which is consistent with our estimate (see the scaling in Eq. \ref{eq v eq} with the velocity $v_{\rm eq}^{4\pi}\simeq6000-7000\,\rm km\,s^{-1}$).
In addition, because of the smaller solid angle the limits on the CNM density become $\simeq7$ times larger than those obtained for a spherical outflow (see also Eq. \ref{eq density limit para}). 

If the assumed effective solid angle is too narrow the implied jet velocity becomes relativistic $v\sim c$. In this case the analysis does not apply as the emission is beamed. 
For ASASSN-14li, this minimal solid angle is roughly $\Omega\gtrsim0.003$ or $\theta\gtrsim0.033\,\rm rad\simeq1.3^\circ$.

Finally, we note that depending on the jet's mass and the external density it may decelerate significantly while propagating in the CNM. If the distribution $M_{\rm ej}(>v)$ and the external density profile are given one can calculate the hydrodynamic evolution and the corresponding emission in a similar manner to those calculated in \S \ref{sec unbound}.

\section{Relativistic jet}\label{sec jet}
A small fraction of TDEs detected as a bright hard X-ray source \citep{Bloom+2011,Burrows+2011,Cenko+2012,Brown+2015} are interpreted as launching relativistic jets, so-called ``jetted TDEs'' \citep[see][for a review]{DeColle&Lu2020}.
Sw1644 is a well-observed prototype of this group.
Its huge isotropic-equivalent X-ray luminosity $\sim10^{47}\,\rm erg\,s^{-1}$ lasting for $\sim10^{6}\,\rm s$ and the radio data strongly suggest that a relativistic jet is launched in this event \citep{Bloom+2011,Burrows+2011,Zauderer+2011,Berger+2012,Zauderer+2013}.

Such jets behave differently from Newtonian ones and we discuss them separately in this section.
Initially a relativistic jet has a narrow opening angle and as its radiation is beamed, only observers within the small opening angle can detect its emission. 
As it sweeps up the CNM and decelerates, the emission becomes less beamed and can be observed from wider angles.
We focus on this late phase during which we can observe the jet regardless of its direction.
Since at early time we observe jetted TDEs only when the jets point toward us, the fraction of jetted TDEs among total number of TDEs is poorly constrained ($\gtrsim3\times10^{-3}$, \citealt{DeColle&Lu2020}).
Radio upper-limits at late time are useful to constrain the event rate as well as the energy of jets pointing away from us that possibly accompany TDEs.

\begin{figure}
\begin{center}
\includegraphics[width=85mm, angle=0]{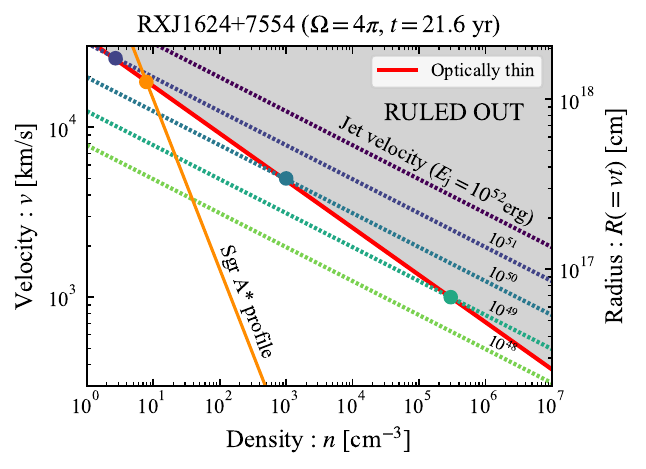}
\caption{
The same as Fig. \ref{fig limit un} but for relativistic jets with different energies.
}
\label{fig limit jet}
\end{center}
\begin{center}
\includegraphics[width=85mm, angle=0]{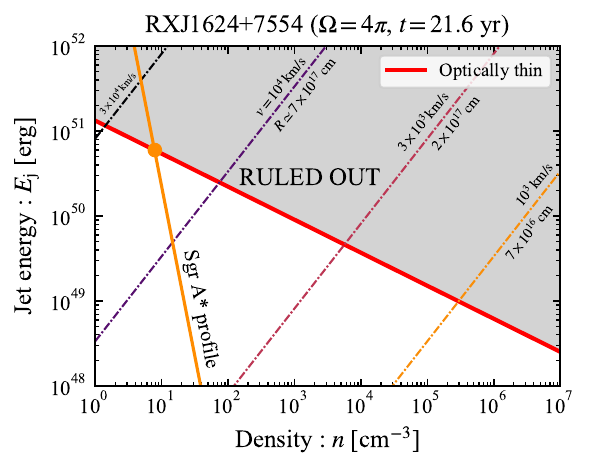}
\caption{The density and jet energy ruled out by the no radio detection in RXJ1624+7554 at $t=21.6\,\rm yr$ for the case of a decelerated jet with $\Omega=4\pi$.
Red line represents the boundary of the excluded region corresponding to the optically thin regime (Eq. \ref{eq jet energy}).
Each colored dash-dotted line shows the contour of each velocity and radius (see Eq. \ref{eq jet velocity}).
The orange line denotes the energy of a jet expanding in Sgr A*-like profile.
The intersection of this trajectory and ruled-out region gives the maximal jet energy for the Sgr A* profile (orange dot).
}
\label{fig nE}
\end{center}
\end{figure}

At late time an initially relativistic jet slows down and becomes Newtonian.
Even if it is not completely spherical its emission would not be beamed.
Energy conservation (Eq. \ref{eq conservation}), enables us to estimate the decelerated jet's velocity at this phase
\begin{align}
v&\simeq\biggl(\frac{2E_{\rm j}}{\Omega\MP n t^3}\biggl)^{1/5}\simeq2.0\times10^5{\,\rm km\,s^{-1}\,}E_{\rm j,51}^{1/5}n_0^{-1/5}t_{\rm yr}^{-3/5}\biggl(\frac{\Omega}{4\pi}\biggl)^{-1/5}\ ,
	\label{eq jet velocity}
\end{align}
where we neglected the jet mass and approximated $R\simeq vt$.
Depending on the initial properties of the jet and the external density profile the decelerating jet may remain non-spherical for a long time \citep[see also Fig. \ref{fig jet image} in Appendix \ref{append accurate}]{Irwin+2019}.
To take this into account we assume that the blast wave subtends a solid angle $\Omega$ into which all the jet's energy is dissipated.
The exact value of $\Omega$ depends  on the details of the hydrodynamic evolution and the sideways propagation of the jet.
However, as we show below, the resulting flux and limit on the energy depend weakly on $\Omega$.
Hence, the exact determination of $\Omega$ is unimportant. 
The fact that the outflow remains jetted for a long time may imply that it also remains relativistic for a longer period, making it easier to hide a powerful jet that is pointing in a different directions even years after it was launched.

An estimate of the energy at which the jet becomes non-relativistic  is given by setting $v\sim c$ in Eq. \eqref{eq jet velocity}:
\begin{align}
E_{\rm j,rel}\simeq\Omega\MP n c^5t^3/2\simeq8.0\times10^{54}{\,\rm erg\,}n_{0}t_{\rm 10 yr}^3\biggl(\frac{\Omega}{4\pi}\biggl).
	\label{eq rela jet}
\end{align}
Hereafter, we change the normalization of time to $t_{10\rm yr}=t/10\,\rm yr$ corresponding to the observed values for the late-time upper-limits.
For the Sgr A* profile, that we use below, the transition to Newtonian takes place at $t \simeq0.34 {\,\rm yr\,} E_{\rm j,52}^{1/2}(\Omega/4\pi)^{-1/2}$.

With Eqs. \eqref{eq velocity limit} and \eqref{eq jet velocity}, the upper limits on the jet energy are given by
\begin{align}
E_{\rm j,51}\lesssim\begin{cases}
\big(0.60\big)_{p=2.5}\,\Big[\bepen^{-1}\varepsilon_{\rm B,-2}^{-\frac{p+1}{4}}t_{\rm 10yr}^{\frac{3(p+1)}{10}}\nu_{\rm 3GHz}^{\frac{p-1}{2}}\\
\,\,\,\,\,\,d_{\rm L,27}^2F_{\rm 30\mu Jy}n_0^{-\frac{3(p+1)}{20}}\big(\frac{\Omega}{4\pi}\big)^{\frac{p+1}{10}}\Big]^{\frac{10}{p+11}}&\text{: $v<v_{\rm DN}$,}\\
\big(0.74\big)_{p=2.5}\,\Big[\bepen^{1-p}\varepsilon_{\rm B,-2}^{-\frac{p+1}{4}}t_{\rm 10yr}^{\frac{3(5p-7)}{10}}\nu_{\rm 3GHz}^{\frac{p-1}{2}}\\
\,\,\,\,\,\,d_{\rm L,27}^2F_{\rm 30\mu Jy}n_0^{\frac{5p-19}{20}}\big(\frac{\Omega}{4\pi}\big)^{\frac{5p-7}{10}}\Big]^{\frac{10}{5p+3}}&\text{: $v>v_{\rm DN}$.}
\end{cases}
    \label{eq jet energy}
\end{align}
Fig. \ref{fig limit jet} depicts trajectories of the decelerated jet's velocity and the constraint from the radio upper-limit of RXJ1624+7554.
We consider only the optically thin regime as at late time the observed frequency is typically above the SSA frequency.
As the limit depends on the velocity, once we obtain a limit on the  jet energy assuming either  $v>$ (or $<$)$v_{\rm DN}$, we have to check using and Eq. \eqref{eq jet velocity} that the velocity satisfies the relevant condition.
Clearly, the limits which we obtain assuming the system in a Newtonian phase, do not hold for extremely energetic jets that are observed early on and still relativistic at the observation epoch.
The critical energy above which our assumption breaks down is given by Eq. \eqref{eq rela jet}.
We miss such relativistic jets if they point away from our line of sight.

Fig. \ref{fig nE} depicts the excluded region in the density and jet energy space for the null detection in RXJ1624+7554.
For each density, the maximal jet energy is given by Eq. \eqref{eq jet energy}.
Assuming Sgr A* like density profile, $n\simeq10{\,\rm cm^{-3}\,}(R/10^{18}\,\rm cm)^{-1}$ we can obtain a limit on the energy of the jets.
Note that once a density profile is given we can calculate the radius and velocity more accurately than Eq. \eqref{eq jet velocity}. However this does not change the results significantly (see Appendix \ref{append accurate}).
The limits for Sgr A* profile are:
\begin{align}
E_{\rm j,51}&\lesssim\begin{cases}
\big(0.21\big)_{p=2.5}\,\Big[\bepen^{-1}\varepsilon_{\rm B,-2}^{-\frac{p+1}{4}}\nu_{\rm 3GHz}^{\frac{p-1}{2}}\\
\,\,\,\,\,\,t_{\rm 10 yr}^{\frac{3(p+1)}{8}}d_{\rm L,27}^2F_{\rm 30\mu Jy}\big(\frac{\Omega}{4\pi}\big)^{\frac{p+1}{16}}\Big]^{\frac{16}{p+17}}&: v<v_{\rm DN}\ ,\\
\big(0.44\big)_{p=2.5}\,\Big[\bepen^{1-p}\varepsilon_{\rm B,-2}^{-\frac{p+1}{4}}\nu_{\rm 3GHz}^{\frac{p-1}{2}}\\
\,\,\,\,\,\,t_{\rm 10yr}^{\frac{11p-13}{8}}d_{\rm L,27}^2F_{\rm 30\mu Jy}\big(\frac{\Omega}{4\pi}\big)^{\frac{3(3p-5)}{16}}\Big]^{\frac{16}{9p+1}}&: v>v_{\rm DN}\ .\\
\end{cases}
    \label{eq jet energy sgr}
\end{align}
Note that the dependence on $\Omega$ is relatively weak, $\propto\Omega^{0.18}$ and $\Omega^{0.32}$ for $p=2.5$.
As shown in Fig. \ref{fig nE}, for RXJ1624+7554 with $F_{\nu}\lesssim50{\,\mu \rm Jy}$ at $t\simeq20{\,\rm yr}$ and Sgr A* like density profile, the energy is constrained to be $E_{\rm j}\lesssim6\times10^{50}\,\rm erg$.
Importantly, with higher densities, as suggested in some radio-detected TDEs, the energy limit will be more severe.

For radio loud TDEs, we can calculate limits on the combinations of the jet energy, $E_{\rm j}$, and CNM density, $n$, as shown in Eq. \eqref{eq jet energy}.
In particular, when the SSA frequency is identified  as a spectral peak $\nu_{\rm p}$, we can directly estimate the energy corresponding to the minimal velocities $v_{\rm eq}$ in Eq. \eqref{eq v eq}.
This energy is given by plunging $v_{\rm eq}$ and $n_{\rm eq}$ into Eq. \eqref{eq jet velocity}:\footnote{Since our fiducial parameter values correspond to the deep-Newtonian phase ($v_{\rm eq}<v_{\rm DN}$), there is a gap in the values of $E_{\rm j,eq}$ for two cases. Additionally these values are much smaller than $E_{\rm j}$ in Eq. \eqref{eq jet energy sgr} because we are using the minimal velocity to derive Eq. \eqref{eq equipartition jet energy}.}
\begin{align}
E_{\rm j,eq}\simeq\begin{cases}
\big(4.2\times10^{47}{\,\rm erg}\big)_{p=2.5}\,\Big[\bepen^{-11}\varepsilon_{\rm B,-2}^{-2(p+1)}\\
\,\,\,F_{30\mu\rm Jy}^{3p+14}d_{\rm L,27}^{2(3p+14)}\big(\frac{\Omega}{4\pi}\big)^{-(p+1)}\Big]^{\frac{1}{2p+13}}\nu_{\rm p,3GHz}^{-1}&: v<v_{\rm DN},\\
\big(1.0\times10^{49}{\,\rm erg}\big)_{p=2.5}\,\Big[\bepen^{11(1-p)}\varepsilon_{\rm B,-2}^{2(1-2p)}\nu_{\rm p,3GHz}^{18p-53}\\
\,\,\,t_{\rm 10yr}^{22(p-2)}F_{30\mu\rm Jy}^{5(6-p)}d_{\rm L,27}^{10(6-p)}\big(\frac{\Omega}{4\pi}\big)^{9p-21}\Big]^{\frac{1}{4p+9}}&: v>v_{\rm DN}.
\end{cases}
	\label{eq equipartition jet energy}
\end{align}
Based on the radio observations, the jet energy of Sw1644 has been estimated by many authors \citep{Berger+2012,Metzger+2012,Zauderer+2013,BarniolDuran&Piran2013,Mimica+2015,Eftekhari+2018,DeColle&Lu2020,Cendes+2021}.
At late time $\gtrsim300\,\rm days$, the outflow decelerates to the Newtonian phase and our method can be applied.
By using Eq. \eqref{eq equipartition jet energy} we estimate the jet energy $E_{\rm j,eq}\sim10^{52}\,\rm erg$, which is consistent with the other estimates at this stage \citep{BarniolDuran&Piran2013,DeColle&Lu2020,Cendes+2021}.

\begin{figure}
\begin{center}
\includegraphics[width=85mm, angle=0]{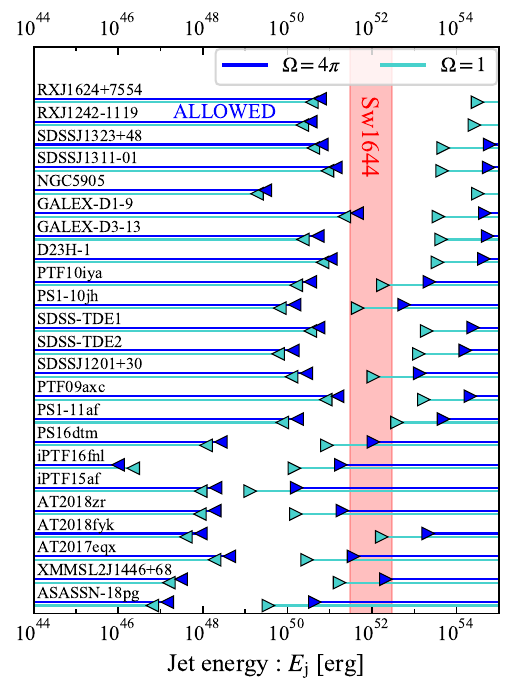}
\caption{Jet energy allowed by radio null detection assuming a CNM density similar to the Sgr A* profile.
At the time of observations the outflow is assumed to have a solid angle $\Omega=4\pi$ (blue) or
$1$ (turquoise - conservatively corresponding to the half-opening angle of $\simeq0.57{\,\rm rad}\simeq33^\circ$, see Fig. \ref{fig jet image}).
Limits obtained more than a few years after the TDE rule out jets with energy $E_{\rm j}\sim 10^{52}\,\rm erg$ similar to Sw1644.
For limits at early time ($\lesssim0.1\,\rm yr$), we cannot reject jets with $E_{\rm j}\gtrsim10^{51-52}$ erg (Eq. \ref{eq rela jet}) because such jets are still relativistic and may beam their radiation away from our line of sight.}
\label{fig Ej}
\end{center}
\end{figure}

For  TDEs with radio upper-limits, we calculate the upper limits on the jet energy assuming an Sgr A* like profile (Eq. \ref{eq jet energy sgr}). 
Fig. \ref{fig Ej} depicts the allowed jet energy for each upper limit (see also Table \ref{table limit}).
At $t \gtrsim $ a few years we can assume that a jet with reasonable energy became Newtonian.
We find that typical upper limits are in the range $10^{50-52}\,\rm erg$ for events with late observations.
Among those, an energetic jet similar to Sw1644 is excluded in events observed more than a few years after the TDE, regardless of the opening angle.
For events with early observations we obtain limits in the range $10^{47-49}\,\rm erg$.
However, in these cases we cannot rule out much more energetic jets that point away from us and are still in the relativistic regime (Eq. \ref{eq rela jet}).
Note that these values are for Sgr A* like density distribution.
The limits will be stronger if the surrounding density is larger, for example like the one inferred for ASASSN-14li \citep{Alexander+2016,Krolik+2016}.

\cite{vanVelzen+2013} constrained the jet energy for several TDEs with radio upper-limits to be $E_{\rm j}<10^{52}\,\rm erg$ by converting the on-axis light curve for Sw1644 \citep{Metzger+2012,Berger+2012} to off-axis ones.
But this estimate ignored the sideways expansion and is invalid for the late phase when the observations were carried out.
\cite{Generozov+2017} derived upper limits on the jet energy based on numerical simulation \citep{Mimica+2015}.
Remarkably, their constraints $E_{\rm j}\lesssim10^{53}\,\rm erg$ are $\simeq20-60$ times weaker than ours for the same parameter values. 
We find that the scaling of their light curves are consistent with ours but the peak time and peak luminosity (when the jet becomes optically thin, $\nu\simeq\nua$) are about 4 and 10 times shorter and dimmer than our estimates, respectively.
In Appendix \ref{append accurate} we provide a specific detailed calculation for the limit on RXJ1624+7554 as an example so that our method can be directly compared to theirs. 
In this case our upper limit of $E_{\rm j}\lesssim7\times10^{50}$ erg is about 30 times smaller than the one obtained by \cite{Generozov+2017}.
Comparison of our hydrodynamics simulations (see Fig. \ref{fig jet image}) to theirs shows consistency, suggesting that the difference originates in the calculations of the synchrotron emission. In spite of some joint efforts we could not identify the origin of this discrepancy.
Note that \cite{Mimica+2015} find a jet energy of  $\simeq5\times10^{53}\,\rm erg$ for Sw1644, which is larger by a similar factor than our and other estimates \citep{BarniolDuran&Piran2013,DeColle&Lu2020,Cendes+2021} of $E_{\rm j}\sim10^{52}\,\rm erg$.

\section{Summary}\label{sec summary}
We analyzed the radio observations and upper limits of TDEs within a model in which the radio emission arises due to the interaction of an outflow with the CNM leading to synchrotron emission from the shocked material.
Within this model we constrained the outflow properties and the CNM density in the galactic nuclear regions.
Our analysis systematically constrains the combination $v \Omega^{a}$, where $a \simeq 0.5$.
Depending on the details of the parameters of the outflow velocity $v$ and the solid angle $\Omega$, it also constrains the CNM density $n$ at the observation epoch.
When the spectral peak is available, our method is similar to the equipartition method \citep{Chevalier1998,BarniolDuran+2013}.
However, we obtained significant limits also in cases in which only a single frequency is observed or in which there are just upper limits. 
In particular, we found that a given detection implies a minimal outflow velocity (Eq. \ref{eq v eq}), that depends only on the solid angle $\Omega$ and on the assumed (or measured) value of the electrons distribution index $p$.
We considered four outflow models: disk wind, unbound debris, and Newtonian jets, corresponding to spherical, wedge and conical geometries, and relativistic jets.

Our calculations considered the 
deep-Newtonian regime ($v<v_{\rm DN} \simeq6\times 10^4\,\rm km\,s^{-1}$) in which only a small fraction $\sim(v/v_{\rm DN})^2\simeq 0.01-0.03$ of the total number of electrons is relativistic and contributes to synchrotron emission. 
This implies that our estimates of the CNM density and swept-up CNM mass are $\sim30-100$ times larger than those found in the previous works based on the equipartition method \citep{Alexander+2016,Krolik+2016,Anderson+2020,Stein+2021,Cendes+2021b} that did not take this effect into account.

Isotropic disk winds with an initial velocity $v_{\rm in}\simeq10^4\,\rm km\,s^{-1}$ can easily produce the observed radio flares as discussed in previous analysis \citep{Alexander+2016,Anderson+2020,Stein+2021,Cendes+2021b}.
For radio-detected TDEs, in particular for events with a measured spectral peak, we found that the wind does not decelerate significantly.
This suggests that the wind's mass should be larger than the swept-up CNM mass $M_{\rm ej}\gtrsim M(R)\simeq0.01-0.1\,\Msun$.
Therefore, current observations do not necessarily require a massive disk wind as suggested by the reprocessed model $M_{\rm ej}\simeq0.5\Msun$ \citep{Metzger&Stone2016}.
Assuming the same typical velocity $v_{\rm in}=10^{4}\,\rm km\,s^{-1}$, we found that some of radio null detection require CNM densities significantly smaller than those required in radio-detected TDEs.
This means that not all TDEs drive sufficiently fast isotropic outflows or the density profiles of galactic nuclear regions vary among galaxies. 
For the TDEs with upper limit determined at ten or more years after the event the implied density upper-limits become comparable to the CNM density around Sgr A*. 

We found that unbound debris with mass $\simeq0.5\,\Msun$, velocity $\sim10^4\,\rm km\,s^{-1}$, and (bow shock's) solid angle $\Omega\simeq0.1$, cannot produce the detectable radio flare due to its too small solid angle.
While this limit is slightly relaxed if we take into account that the shock velocity is larger than the fluid velocity, this is not sufficient to fully resolve the discrepancy.
However, as suggested  by \cite{Yalinewich+2019b}, higher velocities and larger solid angles ($\Omega\sim 1$) can arise when the stellar pericenter is well below the tidal radius (deep penetration). In such cases the debris can produce a detectable radio flare \citep{Krolik+2016}. 
Rarity of such deep-penetration events may explain why radio detection is not common.

We also considered conical outflows (Newtonian jets) and found that those could be the observed radio sources.
Such jets can be realized, for instance, by a disk wind with an angular structure.
Lacking a specific model for the formation of such jets and no definite predictions on their properties, one can always find model parameters that fit the observations.
We can only derive scaling relations between the minimal velocity and the required density for the solid angle $\Omega<4\pi$ (assuming the jet mass is larger than the swept-up CNM mass).
Both velocity and density increase roughly $\propto \Omega^{-0.5}$.

Radio upper-limits also constrain the energy of relativistic jets accompanying TDEs.
We can safely reject jets as energetic as Sw1644, $E_{\rm j}\sim10^{52}\,\rm erg$ for TDEs with upper limits at $t\gtrsim$ a few years.
Note however, that it is impossible to exclude extremely energetic jets with energies $> 10^{53}\,\rm erg$ that have not decelerated yet to the Newtonian phase.
Future late observations will however, rule out or identify such jets. 
For upper limits at $t\lesssim0.01-0.1\,\rm yr$, we have very tight constraints on the existence of jets with energy $E_{\rm j}\lesssim10^{47-49}\,\rm erg$ while we cannot exclude jets with $E_{\rm j}\gtrsim10^{51}\,\rm erg$.
These jets are still relativistic and hence beamed away from us at the time of observations.
They will be ruled out if no radio signals are detected a few years after the discovery.
Therefore we encourage continuous radio monitoring to search for energetic off-axis TDE jets.

Radio-detected TDEs have been suggested as a tool to explore the CNM density profiles of distant galactic centers.
So far we could explore this profile only for the Milky Way. However, our results show that without knowing the outflow's geometry $\Omega$ and the velocity $v$ (or more precisely the velocity distribution $M_{\rm ej}(>v)$), it is in fact impossible to determine the exact value of the CNM profile from the radio observations alone.
With reasonable assumptions we can estimate the profile's radial behavior but not the exact normalization.
Within the models that we considered, the CNM density of radio-detected events is $\gtrsim10^2$ times larger than that of Sgr A* profile,
and its slope is typically $n(R)\propto R^{-3}$.

The total luminosity from synchrotron emission is determined by the peak of $\nu F_{\nu}$.
For slow cooling, which is relevant in all cases considered, this peak corresponds to the cooling frequency \citep[e.g.][]{Sari+1998}:
\begin{align}
\nu_{\rm c}&=\frac{18\pi \ME c e}{\sigma_{\rm T}^2B^3t^2}\simeq6.2\times10^{18}{\,\rm Hz\,}\varepsilon_{\rm B,-2}^{-3/2}n_0^{-3/2}v_9^{-3}t_{\rm yr}^{-2}\ .
\end{align}
As we have seen, for radio-detected events the required density is $n\sim10^{5}\,\rm cm^{-3}$ implying $\nu_{\rm c}\simeq2.0\times10^{11}\,\rm Hz$.
By using the observed luminosity $\nu L_\nu\sim10^{38}\,\rm erg\,s^{-1}$ at $5\,\rm GHz$ and noting that luminosity scales $\nu L_{\nu}\propto \nu^{(3-p)/2}$, we estimate the total luminosity
\begin{align}
L\sim\nu_{\rm c} L_{\nu_{\rm c}}\simeq\big(2\times10^{38}{\,\rm erg\,s^{-1}}\big)_{p=2.5}\,\nu_{\rm c,11}^{\frac{3-p}{2}}\ .
\end{align}
For reasonable parameters, the peak luminosity is less than $10^{40}\,\rm erg\,s^{-1}$.
The total emitted energy $\sim tL\simeq10^{47}\,\rm erg$ is much smaller than the kinetic energy $M_{\rm ej}v^2/2\sim10^{50}{\,\rm erg\,}M_{\rm ej,-1}v_9^2$, for the sources considered here.
This implies that, unlike claims in the literature \citep[e.g.][]{Cendes+2021b,Stein+2021}, to produce the observed emission there is no need to consider additional energy-injection into the outflow (Matsumoto et al. in prep).

To conclude we have shown here the importance of late radio followup of TDEs.
Somewhat surprisingly these observations whose schedule can be flexible are very significant.
Upper limits taken 20 years or more after the event provide interesting limits on properties of the outflow, the nature of the TDE, and on the density structure surrounding the supermassive BH.

\section*{acknowledgments}
We thank Matteo Pais, who conducted the numerical simulations
shown in Appendix \ref{append accurate} and graciously shared his data with us, for his
generous assistance. We also thank Chi-Ho Chan, Julian Krolik, Ben Margalit, and Ehud Nakar for fruitful discussions and helpful comments and Aleksey Generozov, Petar Mimica, and Nicholas Stone for sharing their unpublished results.
This work is supported in part by JSPS Postdoctral Fellowship, Kakenhi No. 19J00214 (T.M.) and by ERC advanced grant ``TReX'' (T.P.).

\section*{data availability}
The data underlying this article will be shared on reasonable request to the corresponding author.

\bibliographystyle{mnras}
\bibliography{reference_matsumoto}

\begin{thebibliography}{}
\makeatletter
\relax
\def\mn@urlcharsother{\let\do\@makeother \do\$\do\&\do\#\do\^\do\_\do\%\do\~}
\def\mn@doi{\begingroup\mn@urlcharsother \@ifnextchar [ {\mn@doi@}
  {\mn@doi@[]}}
\def\mn@doi@[#1]#2{\def\@tempa{#1}\ifx\@tempa\@empty \href
  {http://dx.doi.org/#2} {doi:#2}\else \href {http://dx.doi.org/#2} {#1}\fi
  \endgroup}
\def\mn@eprint#1#2{\mn@eprint@#1:#2::\@nil}
\def\mn@eprint@arXiv#1{\href {http://arxiv.org/abs/#1} {{\tt arXiv:#1}}}
\def\mn@eprint@dblp#1{\href {http://dblp.uni-trier.de/rec/bibtex/#1.xml}
  {dblp:#1}}
\def\mn@eprint@#1:#2:#3:#4\@nil{\def\@tempa {#1}\def\@tempb {#2}\def\@tempc
  {#3}\ifx \@tempc \@empty \let \@tempc \@tempb \let \@tempb \@tempa \fi \ifx
  \@tempb \@empty \def\@tempb {arXiv}\fi \@ifundefined
  {mn@eprint@\@tempb}{\@tempb:\@tempc}{\expandafter \expandafter \csname
  mn@eprint@\@tempb\endcsname \expandafter{\@tempc}}}

\bibitem[\protect\citeauthoryear{{Alexander}, {Berger}, {Guillochon},
  {Zauderer}  \& {Williams}}{{Alexander} et~al.}{2016}]{Alexander+2016}
{Alexander} K.~D.,  {Berger} E.,  {Guillochon} J.,  {Zauderer} B.~A.,
  {Williams} P.~K.~G.,  2016, \mn@doi [\apjl] {10.3847/2041-8205/819/2/L25},
  \href {https://ui.adsabs.harvard.edu/abs/2016ApJ...819L..25A} {819, L25}

\bibitem[\protect\citeauthoryear{{Alexander}, {Wieringa}, {Berger}, {Saxton}
  \& {Komossa}}{{Alexander} et~al.}{2017}]{Alexander+2017}
{Alexander} K.~D.,  {Wieringa} M.~H.,  {Berger} E.,  {Saxton} R.~D.,
  {Komossa} S.,  2017, \mn@doi [\apj] {10.3847/1538-4357/aa6192}, \href
  {https://ui.adsabs.harvard.edu/abs/2017ApJ...837..153A} {837, 153}

\bibitem[\protect\citeauthoryear{{Alexander}, {van Velzen}, {Horesh}  \&
  {Zauderer}}{{Alexander} et~al.}{2020}]{Alexander+2020}
{Alexander} K.~D.,  {van Velzen} S.,  {Horesh} A.,   {Zauderer} B.~A.,  2020,
  \mn@doi [\ssr] {10.1007/s11214-020-00702-w}, \href
  {https://ui.adsabs.harvard.edu/abs/2020SSRv..216...81A} {216, 81}

\bibitem[\protect\citeauthoryear{{Alexander} et~al.,}{{Alexander}
  et~al.}{2021}]{Alexander+2021}
{Alexander} K.~D.,  et~al., 2021, Transient Name Server AstroNote, \href
  {https://ui.adsabs.harvard.edu/abs/2021TNSAN..24....1A} {24, 1}

\bibitem[\protect\citeauthoryear{{Anderson} et~al.,}{{Anderson}
  et~al.}{2020}]{Anderson+2020}
{Anderson} M.~M.,  et~al., 2020, \mn@doi [\apj] {10.3847/1538-4357/abb94b},
  \href {https://ui.adsabs.harvard.edu/abs/2020ApJ...903..116A} {903, 116}

\bibitem[\protect\citeauthoryear{{Arcavi} et~al.,}{{Arcavi}
  et~al.}{2014}]{Arcavi+2014}
{Arcavi} I.,  et~al., 2014, \mn@doi [\apj] {10.1088/0004-637X/793/1/38}, \href
  {https://ui.adsabs.harvard.edu/abs/2014ApJ...793...38A} {793, 38}

\bibitem[\protect\citeauthoryear{{Baganoff} et~al.,}{{Baganoff}
  et~al.}{2003}]{Baganoff+2003}
{Baganoff} F.~K.,  et~al., 2003, \mn@doi [\apj] {10.1086/375145}, \href
  {http://ads.nao.ac.jp/abs/2003ApJ...591..891B} {591, 891}

\bibitem[\protect\citeauthoryear{{Barniol Duran} \& {Piran}}{{Barniol Duran} \&
  {Piran}}{2013}]{BarniolDuran&Piran2013}
{Barniol Duran} R.,  {Piran} T.,  2013, \mn@doi [\apj]
  {10.1088/0004-637X/770/2/146}, \href
  {https://ui.adsabs.harvard.edu/abs/2013ApJ...770..146B} {770, 146}

\bibitem[\protect\citeauthoryear{{Barniol Duran}, {Nakar}  \& {Piran}}{{Barniol
  Duran} et~al.}{2013}]{BarniolDuran+2013}
{Barniol Duran} R.,  {Nakar} E.,   {Piran} T.,  2013, \mn@doi [\apj]
  {10.1088/0004-637X/772/1/78}, \href
  {https://ui.adsabs.harvard.edu/abs/2013ApJ...772...78B} {772, 78}

\bibitem[\protect\citeauthoryear{{Begelman} \& {Cioffi}}{{Begelman} \&
  {Cioffi}}{1989}]{Begelman&Cioffi1989}
{Begelman} M.~C.,  {Cioffi} D.~F.,  1989, \mn@doi [\apjl] {10.1086/185542},
  \href {http://adsabs.harvard.edu/abs/1989ApJ...345L..21B} {345, L21}

\bibitem[\protect\citeauthoryear{{Berger}, {Zauderer}, {Pooley}, {Soderberg},
  {Sari}, {Brunthaler}  \& {Bietenholz}}{{Berger} et~al.}{2012}]{Berger+2012}
{Berger} E.,  {Zauderer} A.,  {Pooley} G.~G.,  {Soderberg} A.~M.,  {Sari} R.,
  {Brunthaler} A.,   {Bietenholz} M.~F.,  2012, \mn@doi [\apj]
  {10.1088/0004-637X/748/1/36}, \href
  {https://ui.adsabs.harvard.edu/abs/2012ApJ...748...36B} {748, 36}

\bibitem[\protect\citeauthoryear{{Blagorodnova} et~al.,}{{Blagorodnova}
  et~al.}{2017}]{Blagorodnova+2017}
{Blagorodnova} N.,  et~al., 2017, \mn@doi [\apj] {10.3847/1538-4357/aa7579},
  \href {https://ui.adsabs.harvard.edu/abs/2017ApJ...844...46B} {844, 46}

\bibitem[\protect\citeauthoryear{{Blagorodnova} et~al.,}{{Blagorodnova}
  et~al.}{2019}]{Blagorodnova+2019}
{Blagorodnova} N.,  et~al., 2019, \mn@doi [\apj] {10.3847/1538-4357/ab04b0},
  \href {https://ui.adsabs.harvard.edu/abs/2019ApJ...873...92B} {873, 92}

\bibitem[\protect\citeauthoryear{{Blanchard} et~al.,}{{Blanchard}
  et~al.}{2017}]{Blanchard+2017}
{Blanchard} P.~K.,  et~al., 2017, \mn@doi [\apj] {10.3847/1538-4357/aa77f7},
  \href {https://ui.adsabs.harvard.edu/abs/2017ApJ...843..106B} {843, 106}

\bibitem[\protect\citeauthoryear{{Blandford} \& {Begelman}}{{Blandford} \&
  {Begelman}}{1999}]{Blandford&Begelman1999}
{Blandford} R.~D.,  {Begelman} M.~C.,  1999, \mn@doi [\mnras]
  {10.1046/j.1365-8711.1999.02358.x}, \href
  {http://adsabs.harvard.edu/abs/1999MNRAS.303L...1B} {303, L1}

\bibitem[\protect\citeauthoryear{{Bloom} et~al.,}{{Bloom}
  et~al.}{2011}]{Bloom+2011}
{Bloom} J.~S.,  et~al., 2011, \mn@doi [Science] {10.1126/science.1207150},
  \href {https://ui.adsabs.harvard.edu/abs/2011Sci...333..203B} {333, 203}

\bibitem[\protect\citeauthoryear{{Bower}}{{Bower}}{2011}]{Bower2011}
{Bower} G.~C.,  2011, \mn@doi [\apjl] {10.1088/2041-8205/732/1/L12}, \href
  {https://ui.adsabs.harvard.edu/abs/2011ApJ...732L..12B} {732, L12}

\bibitem[\protect\citeauthoryear{{Bower}, {Metzger}, {Cenko}, {Silverman}  \&
  {Bloom}}{{Bower} et~al.}{2013}]{Bower+2013}
{Bower} G.~C.,  {Metzger} B.~D.,  {Cenko} S.~B.,  {Silverman} J.~M.,   {Bloom}
  J.~S.,  2013, \mn@doi [\apj] {10.1088/0004-637X/763/2/84}, \href
  {https://ui.adsabs.harvard.edu/abs/2013ApJ...763...84B} {763, 84}

\bibitem[\protect\citeauthoryear{{Bromberg}, {Nakar}, {Piran}  \&
  {Sari}}{{Bromberg} et~al.}{2012}]{Bromberg+2012}
{Bromberg} O.,  {Nakar} E.,  {Piran} T.,   {Sari} R.,  2012, \mn@doi [\apj]
  {10.1088/0004-637X/749/2/110}, \href
  {http://adsabs.harvard.edu/abs/2012ApJ...749..110B} {749, 110}

\bibitem[\protect\citeauthoryear{{Brown}, {Levan}, {Stanway}, {Tanvir},
  {Cenko}, {Berger}, {Chornock}  \& {Cucchiaria}}{{Brown}
  et~al.}{2015}]{Brown+2015}
{Brown} G.~C.,  {Levan} A.~J.,  {Stanway} E.~R.,  {Tanvir} N.~R.,  {Cenko}
  S.~B.,  {Berger} E.,  {Chornock} R.,   {Cucchiaria} A.,  2015, \mn@doi
  [\mnras] {10.1093/mnras/stv1520}, \href
  {https://ui.adsabs.harvard.edu/abs/2015MNRAS.452.4297B} {452, 4297}

\bibitem[\protect\citeauthoryear{{Burrows} et~al.,}{{Burrows}
  et~al.}{2011}]{Burrows+2011}
{Burrows} D.~N.,  et~al., 2011, \mn@doi [\nat] {10.1038/nature10374}, \href
  {https://ui.adsabs.harvard.edu/abs/2011Natur.476..421B} {476, 421}

\bibitem[\protect\citeauthoryear{{Cendes}, {Alexander}, {Berger}, {Eftekhari},
  {Williams}  \& {Chornock}}{{Cendes} et~al.}{2021a}]{Cendes+2021b}
{Cendes} Y.,  {Alexander} K.~D.,  {Berger} E.,  {Eftekhari} T.,  {Williams}
  P.~K.~G.,   {Chornock} R.,  2021a, arXiv e-prints, \href
  {https://ui.adsabs.harvard.edu/abs/2021arXiv210306299C} {p. arXiv:2103.06299}

\bibitem[\protect\citeauthoryear{{Cendes}, {Eftekhari}, {Berger}  \&
  {Polisensky}}{{Cendes} et~al.}{2021b}]{Cendes+2021}
{Cendes} Y.,  {Eftekhari} T.,  {Berger} E.,   {Polisensky} E.,  2021b, \mn@doi
  [\apj] {10.3847/1538-4357/abd323}, \href
  {https://ui.adsabs.harvard.edu/abs/2021ApJ...908..125C} {908, 125}

\bibitem[\protect\citeauthoryear{{Cenko} et~al.,}{{Cenko}
  et~al.}{2012}]{Cenko+2012}
{Cenko} S.~B.,  et~al., 2012, \mn@doi [\apj] {10.1088/0004-637X/753/1/77},
  \href {https://ui.adsabs.harvard.edu/abs/2012ApJ...753...77C} {753, 77}

\bibitem[\protect\citeauthoryear{{Chevalier}}{{Chevalier}}{1998}]{Chevalier1998}
{Chevalier} R.~A.,  1998, \mn@doi [\apj] {10.1086/305676}, \href
  {http://adsabs.harvard.edu/abs/1998ApJ...499..810C} {499, 810}

\bibitem[\protect\citeauthoryear{{Chornock} et~al.,}{{Chornock}
  et~al.}{2014}]{Chornock+2014}
{Chornock} R.,  et~al., 2014, \mn@doi [\apj] {10.1088/0004-637X/780/1/44},
  \href {https://ui.adsabs.harvard.edu/abs/2014ApJ...780...44C} {780, 44}

\bibitem[\protect\citeauthoryear{{De Colle} \& {Lu}}{{De Colle} \&
  {Lu}}{2020}]{DeColle&Lu2020}
{De Colle} F.,  {Lu} W.,  2020, \mn@doi [\nar] {10.1016/j.newar.2020.101538},
  \href {https://ui.adsabs.harvard.edu/abs/2020NewAR..8901538D} {89, 101538}

\bibitem[\protect\citeauthoryear{{Eftekhari}, {Berger}, {Zauderer}, {Margutti}
  \& {Alexander}}{{Eftekhari} et~al.}{2018}]{Eftekhari+2018}
{Eftekhari} T.,  {Berger} E.,  {Zauderer} B.~A.,  {Margutti} R.,   {Alexander}
  K.~D.,  2018, \mn@doi [\apj] {10.3847/1538-4357/aaa8e0}, \href
  {https://ui.adsabs.harvard.edu/abs/2018ApJ...854...86E} {854, 86}

\bibitem[\protect\citeauthoryear{{Generozov}, {Mimica}, {Metzger}, {Stone},
  {Giannios}  \& {Aloy}}{{Generozov} et~al.}{2017}]{Generozov+2017}
{Generozov} A.,  {Mimica} P.,  {Metzger} B.~D.,  {Stone} N.~C.,  {Giannios} D.,
    {Aloy} M.~A.,  2017, \mn@doi [\mnras] {10.1093/mnras/stw2439}, \href
  {https://ui.adsabs.harvard.edu/abs/2017MNRAS.464.2481G} {464, 2481}

\bibitem[\protect\citeauthoryear{{Giannios} \& {Metzger}}{{Giannios} \&
  {Metzger}}{2011}]{Giannios&Metzger2011}
{Giannios} D.,  {Metzger} B.~D.,  2011, \mn@doi [\mnras]
  {10.1111/j.1365-2966.2011.19188.x}, \href
  {https://ui.adsabs.harvard.edu/abs/2011MNRAS.416.2102G} {416, 2102}

\bibitem[\protect\citeauthoryear{{Gillessen} et~al.,}{{Gillessen}
  et~al.}{2019}]{Gillessen+2019}
{Gillessen} S.,  et~al., 2019, \mn@doi [\apj] {10.3847/1538-4357/aaf4f8}, \href
  {https://ui.adsabs.harvard.edu/abs/2019ApJ...871..126G} {871, 126}

\bibitem[\protect\citeauthoryear{{Goodwin} et~al.,}{{Goodwin}
  et~al.}{2021}]{Goodwin+2021}
{Goodwin} A.,  et~al., 2021, The Astronomer's Telegram, \href
  {https://ui.adsabs.harvard.edu/abs/2021ATel14439....1G} {14439, 1}

\bibitem[\protect\citeauthoryear{{Hills}}{{Hills}}{1975}]{Hills1975}
{Hills} J.~G.,  1975, \mn@doi [\nat] {10.1038/254295a0}, \href
  {https://ui.adsabs.harvard.edu/abs/1975Natur.254..295H} {254, 295}

\bibitem[\protect\citeauthoryear{{Holoien} et~al.,}{{Holoien}
  et~al.}{2020}]{Holoien+2020}
{Holoien} T. W.~S.,  et~al., 2020, \mn@doi [\apj] {10.3847/1538-4357/ab9f3d},
  \href {https://ui.adsabs.harvard.edu/abs/2020ApJ...898..161H} {898, 161}

\bibitem[\protect\citeauthoryear{{Horesh}, {Cenko}  \& {Arcavi}}{{Horesh}
  et~al.}{2021}]{Horesh+2021}
{Horesh} A.,  {Cenko} S.~B.,   {Arcavi} I.,  2021, \mn@doi [Nature Astronomy]
  {10.1038/s41550-021-01300-8}, \href
  {https://ui.adsabs.harvard.edu/abs/2021NatAs.tmp...35H} {}

\bibitem[\protect\citeauthoryear{{Huang} \& {Cheng}}{{Huang} \&
  {Cheng}}{2003}]{Huang&Cheng2003}
{Huang} Y.~F.,  {Cheng} K.~S.,  2003, \mn@doi [\mnras]
  {10.1046/j.1365-8711.2003.06430.x}, \href
  {https://ui.adsabs.harvard.edu/abs/2003MNRAS.341..263H} {341, 263}

\bibitem[\protect\citeauthoryear{{Irwin}, {Henriksen}, {Krause}, {Wang},
  {Wiegert}, {Murphy}, {Heald}  \& {Perlman}}{{Irwin}
  et~al.}{2015}]{Irwin+2015}
{Irwin} J.~A.,  {Henriksen} R.~N.,  {Krause} M.,  {Wang} Q.~D.,  {Wiegert} T.,
  {Murphy} E.~J.,  {Heald} G.,   {Perlman} E.,  2015, \mn@doi [\apj]
  {10.1088/0004-637X/809/2/172}, \href
  {https://ui.adsabs.harvard.edu/abs/2015ApJ...809..172I} {809, 172}

\bibitem[\protect\citeauthoryear{{Irwin}, {Nakar}  \& {Piran}}{{Irwin}
  et~al.}{2019}]{Irwin+2019}
{Irwin} C.~M.,  {Nakar} E.,   {Piran} T.,  2019, \mn@doi [\mnras]
  {10.1093/mnras/stz2268}, \href
  {https://ui.adsabs.harvard.edu/abs/2019MNRAS.489.2844I} {489, 2844}

\bibitem[\protect\citeauthoryear{{Komossa}}{{Komossa}}{2002}]{Komossa2002}
{Komossa} S.,  2002, in {Gilfanov} M.,  {Sunyeav} R.,   {Churazov} E.,  eds,
  Lighthouses of the Universe: The Most Luminous Celestial Objects and Their
  Use for Cosmology. p.~436 (\mn@eprint {arXiv} {astro-ph/0109441}),
  \mn@doi{10.1007/10856495_62}

\bibitem[\protect\citeauthoryear{{Komossa}}{{Komossa}}{2015}]{Komossa2015}
{Komossa} S.,  2015, \mn@doi [Journal of High Energy Astrophysics]
  {10.1016/j.jheap.2015.04.006}, \href
  {https://ui.adsabs.harvard.edu/abs/2015JHEAp...7..148K} {7, 148}

\bibitem[\protect\citeauthoryear{{Krolik}, {Piran}, {Svirski}  \&
  {Cheng}}{{Krolik} et~al.}{2016}]{Krolik+2016}
{Krolik} J.,  {Piran} T.,  {Svirski} G.,   {Cheng} R.~M.,  2016, \mn@doi [\apj]
  {10.3847/0004-637X/827/2/127}, \href
  {https://ui.adsabs.harvard.edu/abs/2016ApJ...827..127K} {827, 127}

\bibitem[\protect\citeauthoryear{{Levan} et~al.,}{{Levan}
  et~al.}{2011}]{Levan+2011}
{Levan} A.~J.,  et~al., 2011, \mn@doi [Science] {10.1126/science.1207143},
  \href {https://ui.adsabs.harvard.edu/abs/2011Sci...333..199L} {333, 199}

\bibitem[\protect\citeauthoryear{{Loeb} \& {Ulmer}}{{Loeb} \&
  {Ulmer}}{1997}]{Loeb&Ulmer1997}
{Loeb} A.,  {Ulmer} A.,  1997, \mn@doi [\apj] {10.1086/304814}, \href
  {https://ui.adsabs.harvard.edu/abs/1997ApJ...489..573L} {489, 573}

\bibitem[\protect\citeauthoryear{{Margalit} \& {Piran}}{{Margalit} \&
  {Piran}}{2015}]{Margalit&Piran2015}
{Margalit} B.,  {Piran} T.,  2015, \mn@doi [\mnras] {10.1093/mnras/stv1550},
  \href {https://ui.adsabs.harvard.edu/abs/2015MNRAS.452.3419M} {452, 3419}

\bibitem[\protect\citeauthoryear{{Matsumoto} \& {Piran}}{{Matsumoto} \&
  {Piran}}{2021}]{Matsumoto&Piran2021}
{Matsumoto} T.,  {Piran} T.,  2021, \mn@doi [\mnras] {10.1093/mnras/stab240},
  \href {https://ui.adsabs.harvard.edu/abs/2021MNRAS.502.3385M} {502, 3385}

\bibitem[\protect\citeauthoryear{{Metzger} \& {Stone}}{{Metzger} \&
  {Stone}}{2016}]{Metzger&Stone2016}
{Metzger} B.~D.,  {Stone} N.~C.,  2016, \mn@doi [\mnras]
  {10.1093/mnras/stw1394}, \href
  {https://ui.adsabs.harvard.edu/abs/2016MNRAS.461..948M} {461, 948}

\bibitem[\protect\citeauthoryear{{Metzger}, {Giannios}  \& {Mimica}}{{Metzger}
  et~al.}{2012}]{Metzger+2012}
{Metzger} B.~D.,  {Giannios} D.,   {Mimica} P.,  2012, \mn@doi [\mnras]
  {10.1111/j.1365-2966.2011.20273.x}, \href
  {https://ui.adsabs.harvard.edu/abs/2012MNRAS.420.3528M} {420, 3528}

\bibitem[\protect\citeauthoryear{{Mignone}, {Bodo}, {Massaglia}, {Matsakos},
  {Tesileanu}, {Zanni}  \& {Ferrari}}{{Mignone} et~al.}{2007}]{Mignone+2007}
{Mignone} A.,  {Bodo} G.,  {Massaglia} S.,  {Matsakos} T.,  {Tesileanu} O.,
  {Zanni} C.,   {Ferrari} A.,  2007, \mn@doi [\apjs] {10.1086/513316}, \href
  {https://ui.adsabs.harvard.edu/abs/2007ApJS..170..228M} {170, 228}

\bibitem[\protect\citeauthoryear{{Mimica}, {Giannios}, {Metzger}  \&
  {Aloy}}{{Mimica} et~al.}{2015}]{Mimica+2015}
{Mimica} P.,  {Giannios} D.,  {Metzger} B.~D.,   {Aloy} M.~A.,  2015, \mn@doi
  [\mnras] {10.1093/mnras/stv825}, \href
  {https://ui.adsabs.harvard.edu/abs/2015MNRAS.450.2824M} {450, 2824}

\bibitem[\protect\citeauthoryear{{Nicholl} et~al.,}{{Nicholl}
  et~al.}{2019}]{Nicholl+2019}
{Nicholl} M.,  et~al., 2019, \mn@doi [\mnras] {10.1093/mnras/stz1837}, \href
  {https://ui.adsabs.harvard.edu/abs/2019MNRAS.488.1878N} {488, 1878}

\bibitem[\protect\citeauthoryear{{Piran}, {Nakar}  \& {Rosswog}}{{Piran}
  et~al.}{2013}]{Piran+2013}
{Piran} T.,  {Nakar} E.,   {Rosswog} S.,  2013, \mn@doi [\mnras]
  {10.1093/mnras/stt037}, \href {http://ads.nao.ac.jp/abs/2013MNRAS.430.2121P}
  {430, 2121}

\bibitem[\protect\citeauthoryear{{Rees}}{{Rees}}{1988}]{Rees1988}
{Rees} M.~J.,  1988, \mn@doi [\nat] {10.1038/333523a0}, \href
  {https://ui.adsabs.harvard.edu/abs/1988Natur.333..523R} {333, 523}

\bibitem[\protect\citeauthoryear{{Ricci} et~al.,}{{Ricci}
  et~al.}{2021}]{Ricci+2021}
{Ricci} R.,  et~al., 2021, \mn@doi [\mnras] {10.1093/mnras/staa3241}, \href
  {https://ui.adsabs.harvard.edu/abs/2021MNRAS.500.1708R} {500, 1708}

\bibitem[\protect\citeauthoryear{{Roth}, {Kasen}, {Guillochon}  \&
  {Ramirez-Ruiz}}{{Roth} et~al.}{2016}]{Roth+2016}
{Roth} N.,  {Kasen} D.,  {Guillochon} J.,   {Ramirez-Ruiz} E.,  2016, \mn@doi
  [\apj] {10.3847/0004-637X/827/1/3}, \href
  {https://ui.adsabs.harvard.edu/abs/2016ApJ...827....3R} {827, 3}

\bibitem[\protect\citeauthoryear{{Ryu}, {Krolik}, {Piran}  \& {Noble}}{{Ryu}
  et~al.}{2020}]{Ryu+2020}
{Ryu} T.,  {Krolik} J.,  {Piran} T.,   {Noble} S.~C.,  2020, \mn@doi [\apj]
  {10.3847/1538-4357/abb3cd}, \href
  {https://ui.adsabs.harvard.edu/abs/2020ApJ...904...99R} {904, 99}

\bibitem[\protect\citeauthoryear{{Sari}, {Piran}  \& {Narayan}}{{Sari}
  et~al.}{1998}]{Sari+1998}
{Sari} R.,  {Piran} T.,   {Narayan} R.,  1998, \mn@doi [\apjl]
  {10.1086/311269}, \href {http://adsabs.harvard.edu/abs/1998ApJ...497L..17S}
  {497, L17}

\bibitem[\protect\citeauthoryear{{Saxton}, {Read}, {Esquej}, {Komossa},
  {Dougherty}, {Rodriguez-Pascual}  \& {Barrado}}{{Saxton}
  et~al.}{2012}]{Saxton+2012}
{Saxton} R.~D.,  {Read} A.~M.,  {Esquej} P.,  {Komossa} S.,  {Dougherty} S.,
  {Rodriguez-Pascual} P.,   {Barrado} D.,  2012, \mn@doi [\aap]
  {10.1051/0004-6361/201118367}, \href
  {https://ui.adsabs.harvard.edu/abs/2012A&A...541A.106S} {541, A106}

\bibitem[\protect\citeauthoryear{{Saxton} et~al.,}{{Saxton}
  et~al.}{2019}]{Saxton+2019}
{Saxton} R.~D.,  et~al., 2019, \mn@doi [\aap] {10.1051/0004-6361/201935650},
  \href {https://ui.adsabs.harvard.edu/abs/2019A&A...630A..98S} {630, A98}

\bibitem[\protect\citeauthoryear{{Saxton}, {Komossa}, {Auchettl}  \&
  {Jonker}}{{Saxton} et~al.}{2020}]{Saxton+2020}
{Saxton} R.,  {Komossa} S.,  {Auchettl} K.,   {Jonker} P.~G.,  2020, \mn@doi
  [\ssr] {10.1007/s11214-020-00708-4}, \href
  {https://ui.adsabs.harvard.edu/abs/2020SSRv..216...85S} {216, 85}

\bibitem[\protect\citeauthoryear{{Sironi} \& {Giannios}}{{Sironi} \&
  {Giannios}}{2013}]{Sironi&Giannios2013}
{Sironi} L.,  {Giannios} D.,  2013, \mn@doi [\apj]
  {10.1088/0004-637X/778/2/107}, \href
  {https://ui.adsabs.harvard.edu/abs/2013ApJ...778..107S} {778, 107}

\bibitem[\protect\citeauthoryear{{Stein} et~al.,}{{Stein}
  et~al.}{2021}]{Stein+2021}
{Stein} R.,  et~al., 2021, \mn@doi [Nature Astronomy]
  {10.1038/s41550-020-01295-8}, \href
  {https://ui.adsabs.harvard.edu/abs/2021NatAs...5..510S} {5, 510}

\bibitem[\protect\citeauthoryear{{Strubbe} \& {Quataert}}{{Strubbe} \&
  {Quataert}}{2009}]{Strubbe&Quataert2009}
{Strubbe} L.~E.,  {Quataert} E.,  2009, \mn@doi [\mnras]
  {10.1111/j.1365-2966.2009.15599.x}, \href
  {https://ui.adsabs.harvard.edu/abs/2009MNRAS.400.2070S} {400, 2070}

\bibitem[\protect\citeauthoryear{{Uno} \& {Maeda}}{{Uno} \&
  {Maeda}}{2020}]{Uno&Maeda2020b}
{Uno} K.,  {Maeda} K.,  2020, \mn@doi [\apjl] {10.3847/2041-8213/abca32}, \href
  {https://ui.adsabs.harvard.edu/abs/2020ApJ...905L...5U} {905, L5}

\bibitem[\protect\citeauthoryear{{Wevers} et~al.,}{{Wevers}
  et~al.}{2019}]{Wevers+2019}
{Wevers} T.,  et~al., 2019, \mn@doi [\mnras] {10.1093/mnras/stz1976}, \href
  {https://ui.adsabs.harvard.edu/abs/2019MNRAS.488.4816W} {488, 4816}

\bibitem[\protect\citeauthoryear{{Wevers} et~al.,}{{Wevers}
  et~al.}{2021}]{Wevers+2021}
{Wevers} T.,  et~al., 2021, arXiv e-prints, \href
  {https://ui.adsabs.harvard.edu/abs/2021arXiv210104692W} {p. arXiv:2101.04692}

\bibitem[\protect\citeauthoryear{{Yalinewich}, {Steinberg}, {Piran}  \&
  {Krolik}}{{Yalinewich} et~al.}{2019}]{Yalinewich+2019b}
{Yalinewich} A.,  {Steinberg} E.,  {Piran} T.,   {Krolik} J.~H.,  2019, \mn@doi
  [\mnras] {10.1093/mnras/stz1567}, \href
  {https://ui.adsabs.harvard.edu/abs/2019MNRAS.487.4083Y} {487, 4083}

\bibitem[\protect\citeauthoryear{{Zauderer} et~al.,}{{Zauderer}
  et~al.}{2011}]{Zauderer+2011}
{Zauderer} B.~A.,  et~al., 2011, \mn@doi [\nat] {10.1038/nature10366}, \href
  {https://ui.adsabs.harvard.edu/abs/2011Natur.476..425Z} {476, 425}

\bibitem[\protect\citeauthoryear{{Zauderer}, {Berger}, {Margutti}, {Pooley},
  {Sari}, {Soderberg}, {Brunthaler}  \& {Bietenholz}}{{Zauderer}
  et~al.}{2013}]{Zauderer+2013}
{Zauderer} B.~A.,  {Berger} E.,  {Margutti} R.,  {Pooley} G.~G.,  {Sari} R.,
  {Soderberg} A.~M.,  {Brunthaler} A.,   {Bietenholz} M.~F.,  2013, \mn@doi
  [\apj] {10.1088/0004-637X/767/2/152}, \href
  {https://ui.adsabs.harvard.edu/abs/2013ApJ...767..152Z} {767, 152}

\bibitem[\protect\citeauthoryear{{van Velzen} et~al.,}{{van Velzen}
  et~al.}{2011}]{vanVelzen+2011}
{van Velzen} S.,  et~al., 2011, \mn@doi [\apj] {10.1088/0004-637X/741/2/73},
  \href {https://ui.adsabs.harvard.edu/abs/2011ApJ...741...73V} {741, 73}

\bibitem[\protect\citeauthoryear{{van Velzen}, {Frail}, {K{\"o}rding}  \&
  {Falcke}}{{van Velzen} et~al.}{2013}]{vanVelzen+2013}
{van Velzen} S.,  {Frail} D.~A.,  {K{\"o}rding} E.,   {Falcke} H.,  2013,
  \mn@doi [\aap] {10.1051/0004-6361/201220426}, \href
  {https://ui.adsabs.harvard.edu/abs/2013A&A...552A...5V} {552, A5}

\bibitem[\protect\citeauthoryear{{van Velzen} et~al.,}{{van Velzen}
  et~al.}{2016}]{vanVelzen+2016}
{van Velzen} S.,  et~al., 2016, \mn@doi [Science] {10.1126/science.aad1182},
  \href {https://ui.adsabs.harvard.edu/abs/2016Sci...351...62V} {351, 62}

\bibitem[\protect\citeauthoryear{{van Velzen} et~al.,}{{van Velzen}
  et~al.}{2019}]{vanVelzen+2019}
{van Velzen} S.,  et~al., 2019, \mn@doi [\apj] {10.3847/1538-4357/aafe0c},
  \href {https://ui.adsabs.harvard.edu/abs/2019ApJ...872..198V} {872, 198}

\bibitem[\protect\citeauthoryear{{van Velzen}, {Holoien}, {Onori}, {Hung}  \&
  {Arcavi}}{{van Velzen} et~al.}{2020}]{vanVelzen+2020}
{van Velzen} S.,  {Holoien} T. W.~S.,  {Onori} F.,  {Hung} T.,   {Arcavi} I.,
  2020, \mn@doi [\ssr] {10.1007/s11214-020-00753-z}, \href
  {https://ui.adsabs.harvard.edu/abs/2020SSRv..216..124V} {216, 124}

\makeatother
\end{thebibliography}

\appendix
\section{Formulae for a general power-law index of electron distribution}\label{append p}
We present the formulae for general $p$.
The equations for SSA frequency and fluxes corresponding to Eqs. \eqref{eq nu_a}, \eqref{eq f nu_a}, \eqref{eq f nu1} and \eqref{eq f nu2} are given by
\begin{align}
\nua&\simeq\begin{cases}
7.2\times10^{3}\big[8.2\times10^7(p-1)\,3^{\frac{p+1}{2}}\big]^{\frac{2}{p+4}}{\,\rm Hz\,}\\
\,\,\,\,\,\,\bepen^{\frac{2}{p+4}}\varepsilon_{\rm B,-2}^{\frac{p+2}{2(p+4)}}n_0^{\frac{p+6}{2(p+4)}}v_9^{\frac{p+6}{p+4}}R_{17}^{\frac{2}{p+4}}&:\,v<v_{\rm DN},\\
4.7\big[3.0\times10^{17}(p-1)\,3^{\frac{p+1}{2}}\big]^{\frac{2}{p+4}}{\,\rm Hz\,}\\
\,\,\,\,\,\,\bepen^{\frac{2(p-1)}{p+4}}\varepsilon_{\rm B,-2}^{\frac{p+2}{2(p+4)}}n_0^{\frac{p+6}{2(p+4)}}v_{9}^{\frac{5p-2}{p+4}}R_{17}^{\frac{2}{p+4}}&:\,v_{\rm DN}<v.
\end{cases}
    \label{eq append nua}\\
F_{\nua}&\simeq\begin{cases}
0.62\big[8.2\times10^7(p-1)\,3^{\frac{p+1}{2}}\big]^{\frac{1-p}{p+4}}{\,\rm \mu Jy\,}\bepen^{\frac{5}{p+4}}\\
\,\,\,\varepsilon_{\rm B,-2}^{\frac{2p+3}{2(p+4)}}n_0^{\frac{2p+13}{2(p+4)}}v_9^{\frac{2p+13}{p+4}}R_{17}^{\frac{2p+13}{p+4}}\big(\frac{\Omega}{4\pi}\big)d_{\rm L,27}^{-2}&:\,v<v_{\rm DN},\\
25\big[3.0\times10^{17}(p-1)\,3^{\frac{p+1}{2}}\big]^{\frac{1-p}{p+4}}{\,\rm \mu Jy\,}\bepen^{\frac{5(p-1)}{p+4}}\\
\,\,\,\varepsilon_{\rm B,-2}^{\frac{2p+3}{2(p+4)}}n_0^{\frac{2p+13}{2(p+4)}}v_{9}^{\frac{12p-7}{p+4}}R_{17}^{\frac{2p+13}{p+4}}\big(\frac{\Omega}{4\pi}\big)d_{\rm L,27}^{-2}&:\,v_{\rm DN}<v,
\end{cases}\\
F_{\nu>\nua}&\simeq\begin{cases}
0.62\big[4.1\times10^5\big]^{\frac{1-p}{2}}{\,\rm \mu Jy\,}\bepen\\
\,\,\,\varepsilon_{\rm B,-2}^{\frac{p+1}{4}}n_0^{\frac{p+5}{4}}v_9^{\frac{p+5}{2}}R_{17}^3\big(\frac{\Omega}{4\pi}\big)\nu_{\rm 3GHz}^{\frac{1-p}{2}}d_{\rm L,27}^{-2}&:\,v<v_{\rm DN},\\
25\big[6.4\times10^8\big]^{\frac{1-p}{2}}{\,\rm \mu Jy\,}\bepen^{p-1}\\
\,\,\,\varepsilon_{\rm B,-2}^{\frac{p+1}{4}}n_0^{\frac{p+5}{4}}v_{9}^{\frac{5p-3}{2}}R_{17}^3\big(\frac{\Omega}{4\pi}\big)\nu_{\rm GHz}^{\frac{1-p}{2}}d_{\rm L,27}^{-2}&:\,v_{\rm DN}<v,
\end{cases}
    \label{eq append flux1}\\
F_{\nu<\nua}&\simeq\frac{8.4\times10^4}{(p-1)3^{\frac{p+1}{2}}}{\,\rm \mu Jy\,}\varepsilon_{\rm B,-2}^{-1/4}
    \nonumber\\
&\,\,\,\,\,\,\,\,\,\,\,\,n_0^{-1/4}v_9^{-1/2}R_{17}^2\biggl(\frac{\Omega}{4\pi}\biggl)\nu_{\rm 3GHz}^{5/2}d_{\rm L,27}^{-2},
\end{align}
respectively.
Constraints on the combinations of the density, velocity, and solid angle, are given by 
\begin{align}
n_{0}^{\frac{p+5}{4}}v_9^{\frac{p+11}{2}}\Omega\lesssim&1.9\times10^4\big[4.1\times10^{5}\big]^{\frac{p-1}{2}}
    \nonumber\\
&\,\,\,\,\,\,\bepen^{-1}\varepsilon_{\rm B,-2}^{-\frac{p+1}{4}}t_{\rm yr}^{-3}\nu_{\rm 3GHz}^{\frac{p-1}{2}}d_{\rm L,27}^{2}F_{\rm 30\mu Jy},
\end{align}
for optically thin and the deep-Newtonian phase (Eq. \ref{eq limit f dn}), 
\begin{align}
n_0^{\frac{p+5}{4}}v_{9}^{\frac{5p+3}{2}}\Omega \lesssim& 500\big[6.4\times10^{8}\big]^{\frac{p-1}{2}}
    \nonumber\\
&\,\,\,\,\,\,\bepen^{1-p}\varepsilon_{\rm B,-2}^{-\frac{p+1}{4}}t_{\rm yr}^{-3}\nu_{\rm 3GHz}^{\frac{p-1}{2}}d_{\rm L,27}^{2}F_{\rm 30\mu Jy},
\end{align}
for optically thin and $v>v_{\rm DN}$ case (Eq. \ref{eq limit f}), and
\begin{align}
&n_0^{-1/4}v_9^{3/2}\Omega \lesssim4.5\times10^{-3}(p-1)3^{\frac{p+1}{2}}\varepsilon_{\rm B,-2}^{1/4}t_{\rm yr}^{-2}\nu_{\rm 3GHz}^{-5/2}d_{\rm L,27}^{2}F_{\rm 30\mu Jy}.
\end{align}
for optically thick case (Eq. \ref{eq limit g}).
The limit on velocity corresponding to Eq. \eqref{eq velocity limit} is
\begin{align}
v_9\lesssim\begin{cases}
\Big[1.9\times10^4\big(4.1\times10^{5}\big)^{\frac{p-1}{2}}\bepen^{-1}\varepsilon_{\rm B,-1}^{-\frac{p+1}{4}}\\
\,\,\,t_{\rm yr}^{-3}\nu_{\rm 3GHz}^{\frac{p-1}{2}}d_{\rm L,27}^2F_{\rm 30\mu Jy}n_0^{-\frac{p+5}{4}}\Omega^{-1}\Big]^{\frac{2}{p+11}}&: {\rm thin\,}(v<v_{\rm DN}),\\
\Big[500\big(6.4\times10^{8}\big)^{\frac{p-1}{2}}\bepen^{1-p}\varepsilon_{\rm B,-2}^{-\frac{p+1}{4}}\\
\,\,\,t_{\rm yr}^{-3}\nu_{\rm 3GHz}^{\frac{p-1}{2}}d_{\rm L,27}^{2}F_{\rm 30\mu Jy}n_0^{-\frac{p+5}{4}}\Omega^{-1}\Big]^{\frac{2}{5p+3}}&: {\rm thin\,}(v>v_{\rm DN}),\\
2.7\times10^{-2}(p-1)^{2/3}3^{\frac{p+1}{3}}\varepsilon_{\rm B,-2}^{1/6}\\
\,\,\,t_{\rm yr}^{-4/3}\nu_{\rm 3GHz}^{-5/3}d_{\rm L,27}^{4/3}F_{\rm 30\mu Jy}^{2/3}n_0^{1/6}\Omega^{2/3}&: {\rm thick}.
\end{cases}
\end{align}

The minimal velocities and corresponding densities are given by
\begin{align}
v_{\rm eq, 9}&\simeq\begin{cases}
\big[1.9\times10^4(4.1\times10^5)^{\frac{p-1}{2}}\big]^{\frac{1}{2p+13}}\\
\,\,\,\big[4.5\times10^{-3}(p-1)3^{\frac{p+1}{2}}\big]^{\frac{p+5}{2p+13}}\bepen^{-\frac{1}{2p+13}}\\
\,\,\,\,\,\,\varepsilon_{\rm B,-2}^{\frac{1}{2p+13}} t_{\rm yr}^{-1}\nu_{\rm 3GHz}^{-1}d_{\rm L,27}^{\frac{2(p+6)}{2p+13}}F_{\rm 30\mu Jy}^{\frac{p+6}{2p+13}}\Omega^{-\frac{p+6}{2p+13}}&:v<v_{\rm DN},\\
\big[500(6.4\times10^{8})^{\frac{p-1}{2}}\big]^{\frac{1}{4p+9}}\\
\,\,\,\big[4.5\times10^{-3}(p-1)3^{\frac{p+1}{2}}\big]^{\frac{p+5}{4p+9}}\bepen^{\frac{1-p}{4p+9}}\\
\,\,\,\,\,\,\varepsilon_{\rm B,-2}^{\frac{1}{4p+9}} t_{\rm yr}^{-\frac{2p+13}{4p+9}}\nu_{\rm 3GHz}^{-\frac{2p+13}{4p+9}}d_{\rm L,27}^{\frac{2(p+6)}{4p+9}}F_{\rm 30\mu Jy}^{\frac{p+6}{4p+9}}\Omega^{-\frac{p+6}{4p+9}}&:v>v_{\rm DN},
\end{cases}\\
n_{\rm eq,0}&\simeq\begin{cases}
\big[1.9\times10^4(4.1\times10^5)^{\frac{p-1}{2}}\big]^{\frac{6}{2p+13}}\\
\,\,\,\big[4.5\times10^{-3}(p-1)3^{\frac{p+1}{2}}\big]^{-\frac{2(p+11)}{2p+13}}\bepen^{-\frac{6}{2p+13}}\\
\,\,\,\,\,\,\varepsilon_{\rm B,-2}^{-\frac{2p+7}{2p+13}} t_{\rm yr}^2\nu_{\rm 3GHz}^4d_{\rm L,27}^{-\frac{4(p+8)}{2p+13}}F_{\rm 30\mu Jy}^{-\frac{2(p+8)}{2p+13}}\Omega^{\frac{2(p+8)}{2p+13}}&:v<v_{\rm DN},\\
\big[500(6.4\times10^{8})^{\frac{p-1}{2}}\big]^{\frac{6}{4p+9}}\\
\,\,\,\big[4.5\times10^{-3}(p-1)3^{\frac{p+1}{2}}\big]^{-\frac{2(5p+3)}{4p+9}}\bepen^{\frac{6(1-p)}{4p+9}}\\
\,\,\,\,\,\,\varepsilon_{\rm B,-2}^{-\frac{4p+3}{4p+9}} t_{\rm yr}^{\frac{2(10p-3)}{4p+9}}\nu_{\rm 3GHz}^{\frac{4(7p+3)}{4p+9}}d_{\rm L,27}^{-\frac{20p}{4p+9}}F_{\rm 30\mu Jy}^{-\frac{10p}{4p+9}}\Omega^{\frac{10p}{4p+9}}&:v>v_{\rm DN}.\\
\end{cases}
\end{align}
corresponding to Eqs. \eqref{eq v eq} and \eqref{eq n eq}, respectively.

Constraints on the jet energy for each case (Eq. \ref{eq jet energy}) and minimal jet energies (Eq. \ref{eq equipartition jet energy}) are given by
\begin{align}
E_{\rm j,51}\lesssim\begin{cases}
\Big[1.8\times10^{-7}(150)^p\bepen^{-1}\varepsilon_{\rm B,-2}^{-\frac{p+1}{4}}t_{\rm yr}^{\frac{3(p+1)}{10}}\\
\,\,\,\nu_{\rm 3GHz}^{\frac{p-1}{2}}d_{\rm L,27}^2F_{\rm 30\mu Jy}n_0^{-\frac{3(p+1)}{20}}\big(\frac{\Omega}{4\pi}\big)^{\frac{p+1}{10}}\Big]^{\frac{10}{p+11}}&\text{: $v<v_{\rm DN}$,}\\
\Big[1.8\times10^{-5}(15)^p\bepen^{1-p}\varepsilon_{\rm B,-2}^{-\frac{p+1}{4}}t_{\rm yr}^{\frac{3(5p-7)}{10}}\\
\,\,\,\nu_{\rm 3GHz}^{\frac{p-1}{2}}d_{\rm L,27}^2F_{\rm 30\mu Jy}n_0^{\frac{5p-19}{20}}\big(\frac{\Omega}{4\pi}\big)^{\frac{5p-7}{10}}\Big]^{\frac{10}{5p+3}}&\text{: $v>v_{\rm DN}$,}\\
\end{cases}
\end{align}
\begin{align}
E_{\rm j,eq}\simeq\begin{cases}
1.1\times10^{44}{\,\rm erg\,}\Big[5.8(9.0\times10^{20})^p\\
\,\,\,(5.2)^{p^2}(p-1)^{3(p+1)}\bepen^{-11}\varepsilon_{\rm B,-2}^{-2(p+1)}\\
\,\,\,\,\,\,d_{\rm L,27}^{2(3p+14)}F_{30\mu\rm Jy}^{3p+14}\big(\frac{\Omega}{4\pi}\big)^{-(p+1)}\Big]^{\frac{1}{2p+13}}\nu_{\rm 3GHz}^{-1}&:v<v_{\rm DN},\\
3.3\times10^{44}{\,\rm erg\,}\Big[1.8\times10^{-92}(9.8\times10^{68})^p\\
\,\,\,(0.064)^{p^2}(p-1)^{19-5p}\bepen^{11(1-p)}\varepsilon_{\rm B,-2}^{2(1-2p)}\\
\,\,\,\,\,\,t_{\rm yr}^{22(p-2)}\nu_{\rm 3GHz}^{18p-53}d_{\rm L,27}^{10(6-p)}F_{30\mu\rm Jy}^{5(6-p)}\big(\frac{\Omega}{4\pi}\big)^{9p-21}\Big]^{\frac{1}{4p+9}}&:v>v_{\rm DN}.\\
\end{cases}
\end{align}
For Sgr A* like density profile, the jet energy is constrained by
\begin{align}
E_{\rm j,51}&\lesssim\begin{cases}\Big[1.1\times10^{-7}(86)^p\bepen^{-1}\varepsilon_{\rm B,-2}^{-\frac{p+1}{4}}t_{\rm yr}^{\frac{3(p+1)}{8}}\\
\,\,\,\,\,\,\nu_{\rm 3GHz}^{\frac{p-1}{2}}d_{\rm L,27}^2F_{\rm 30\mu Jy}\big(\frac{\Omega}{4\pi}\big)^{\frac{p+1}{16}}\Big]^{\frac{16}{p+17}}&: v<v_{\rm DN},\\
\Big[6.6\times10^{-7}(35)^p\bepen^{1-p}\varepsilon_{\rm B,-2}^{-\frac{p+1}{4}}t_{\rm yr}^{\frac{11p-13}{8}}\\
\,\,\,\,\,\,\nu_{\rm 3GHz}^{\frac{p-1}{2}}d_{\rm L,27}^2F_{\rm 30\mu Jy}\big(\frac{\Omega}{4\pi}\big)^{\frac{3(3p-5)}{16}}\Big]^{\frac{16}{9p+1}}&: v>v_{\rm DN}.\\
\end{cases}
\end{align}

\section{More precise estimation of the jet energy}\label{append accurate}
\begin{figure}
\begin{center}
\includegraphics[width=120mm, angle=90]{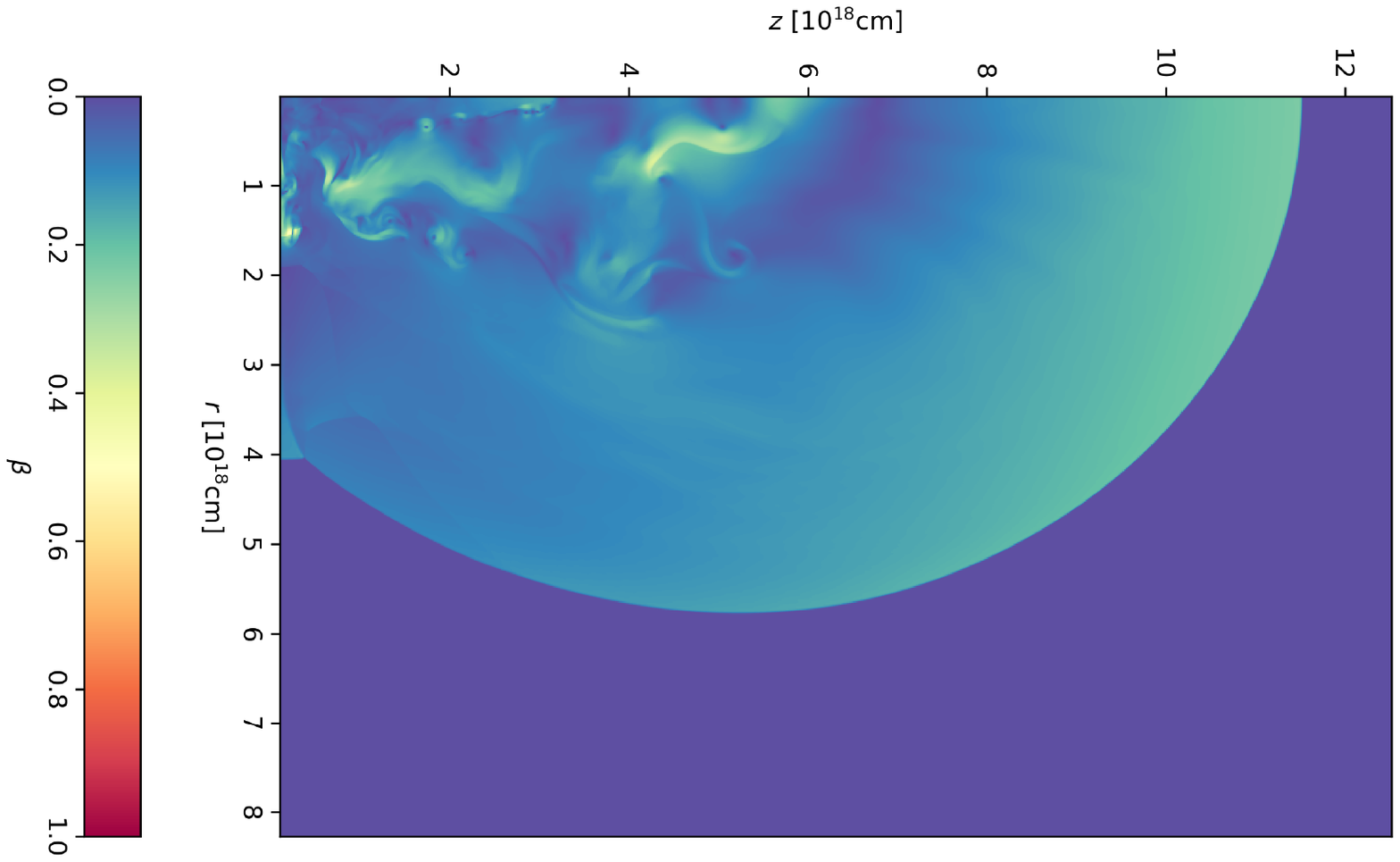}
\caption{2D hydrodynamics simulation of a relativistic jet carried out by Matteo Pais using \texttt{PLUTO} \citep{Mignone+2007}. The jet is injected with energy $E_{\rm j}=10^{53}\,\rm erg$  and half-opening angle $\theta=20^\circ$ into a Sgr A* like CNM, $n(R)=10{\,\rm cm^{-3}\,}(R/10^{18}\,{\rm cm})^{-1}$. 
The velocity ($\beta\equiv{v/c}$) distribution at $t=21.7{\,\rm yr}$ after the launch (the same as the observation of RXJ1624+7554) shows that while the system is almost Newtonian (the highest $\beta$ values are around 0.5) the jet is  not spherical yet. The resulting cocoon has a bipolar shape with an opening angle of $\theta\simeq30^\circ$.
}
\label{fig jet image}
\end{center}
\end{figure}

We present a detailed calculation of the upper limit on the jet energy for RXJ1624+7554.
The radio upper-limit is $F_{\nu}=51\,\rm \mu Jy$ at $\nu=3\,\rm GHz$ with $d_{\rm L}=270\,\rm Mpc$ and $t=21.7\,\rm yr$ (see Table \ref{table limit}).
We consider a Milky-Way like density profile, $n(R)=\tilde{n}(R/\tilde{R})^{-k}$ with $\tilde{n}=10\,\rm cm^{-3}$, $\tilde{R}=10^{18}\,\rm cm$, and $k=1$.
In the Newtonian phase, the jet energy is given by
\begin{align}
E_j=\Omega \MP \frac{v^2}{2}\int_0^R r^2n(r)dr=\frac{\Omega \MP}{2(3-k)}\tilde{n}\tilde{R}^kv^2R^{3-k}.
    \label{eq jet energy append}
\end{align}
Note that in this section we take into account the numerical factors that are ignored in \S \ref{sec jet}.
Using $R\propto t^l$ with $l=2/(5-k)=0.5$, we obtain the relation between the velocity and radius as $v=\frac{dR}{dt}=lR/t$.
Substituting this into Eq. \eqref{eq jet energy append} and setting $v=c$, we obtain the timescale at which the jet becomes Newtonian
\begin{align}
t_{\rm NR}&=\biggl(\frac{2(3-k)l^{3-k}E_{\rm j}}{\Omega \MP c^{5-k}\tilde{n}\tilde{R}^k}\biggl)^{\frac{1}{3-k}}\simeq7.7\times10^{-2}{\,\rm yr\,}E_{\rm j,51}^{1/2}\biggl(\frac{\Omega}{4\pi}\biggl)^{-1/2}.
\end{align}
The jet is Newtonian at the observation time as long as $E_{\rm j}\lesssim E_{\rm j,rel}\simeq8\times10^{55}\,\rm erg\,(\Omega/4\pi)$.
The time evolution of the jet velocity and radius for $t>t_{\rm NR}$ are given by 
\begin{align}
v&=c\biggl(\frac{t}{t_{\rm NR}}\biggl)^{l-1}\simeq18000{\,\rm km\,s^{-1}\,}E_{\rm j,51}^{1/4}\biggl(\frac{t}{21.7{\,\rm yr}}\biggl)^{-1/2}\biggl(\frac{\Omega}{4\pi}\biggl)^{-1/4},\\
R&=\frac{ct_{\rm NR}}{l}\biggl(\frac{t}{t_{\rm NR}}\biggl)^{l}\simeq2.5\times10^{18}{\,\rm cm\,}E_{\rm j,51}^{1/4}\biggl(\frac{t}{21.7{\,\rm yr}}\biggl)^{1/2}\biggl(\frac{\Omega}{4\pi}\biggl)^{-1/4}.
\end{align}
These estimates are reasonably consistent with the 2D jet simulation shown in Fig. \ref{fig jet image} but also with the 1D simulation of \cite{Mimica+2015}.
By using these equations with Eq. \eqref{eq append flux1}, we derive the flux at the optically thin regime:
\begin{align}
F_{\nu}&\simeq\begin{cases}
79{\,\rm\mu Jy\,}\varepsilon_{\rm e,-1}\big(\frac{\varepsilon_{\rm B}}{0.002}\big)^{0.83}E_{\rm j,51}^{1.21}\big(\frac{t}{21.7\,\rm yr}\big)^{-1.24}
    \\
\,\,\,\,\,\,\nu_{\rm 3GHz}^{-0.65}\big(\frac{d_{\rm L}}{270\,\rm Mpc}\big)^{-2}\big(\frac{\Omega}{4\pi}\big)^{-0.21}&: v<v_{\rm DN},\\
36{\,\rm\mu Jy\,}\varepsilon_{\rm e,-1}^{1.3}\big(\frac{\varepsilon_{\rm B}}{0.002}\big)^{0.83}E_{\rm j,51}^{1.36}\big(\frac{t}{21.7\,\rm yr}\big)^{-1.54}
    \\
\,\,\,\,\,\,\nu_{\rm 3GHz}^{-0.65}\big(\frac{d_{\rm L}}{270\,\rm Mpc}\big)^{-2}\big(\frac{\Omega}{4\pi}\big)^{-0.36}&: v>v_{\rm DN},\\
\end{cases}
    \label{eq append flux}
\end{align}
where we adopt the same micro-physics parameters as \cite{Generozov+2017} ($\varepsilon_{\rm e}=0.1$, $\varepsilon_{\rm B}=0.002$, and $p=2.3$).
We find that for RXJ1624+7554, the deep-Newtonian phase ($v<v_{\rm DN}$) is relevant and we obtain a limit of $E_{\rm j}\lesssim7.0\times10^{50}\,\rm erg$, which is $\simeq30$ times smaller than the one of \cite{Generozov+2017}.
It should be noted that \cite{Generozov+2017} did not consider the deep-Newtonian phase and calculated the limit for the normal phase ($v>v_{\rm DN}$).
However, this does not change the limit on jet significantly ($E_{\rm j}\lesssim1.3\times10^{51}\,\rm erg$).

\section{Constraint for each event}
We present our complete results on radio upper-limits (Table \ref{table limit}) and radio-detected events (Table \ref{table detection}).

\begin{landscape}
\begin{table}
\caption{Constraints imposed by radio upper-limits for the cases of isotropic disk wind, unbound debris, and relativistic jet. For the former two cases, we show the minimal velocity $v_{\rm eq}$, corresponding density $n_{\rm eq}$, limiting velocity in the optically thin regime $v_-$, and the corresponding density upper-limit $n_-$.
For relativistic jet, the upper limit $E_{\rm j}$ is relevant up to the energy $E_{\rm j,rel}$.}
\label{table limit}
\begin{tabular}{llrrr|rrrr|rrrr|rr}
\hline
Event (Ref)&Redshift&Time&Freq.&Flux&\multicolumn{4}{c|}{Disk wind ($\Omega=4\pi$)}&\multicolumn{4}{c|}{Unbound debris ($\Omega=0.1$)}&\multicolumn{2}{c}{Relativistic jet ($\Omega=4\pi$)}\\
&$z$&$t$&$\nu$&$F_{\nu}$&$v_{\rm eq}$&$n_{\rm eq}$&$v_{-}$&$n_{-}$&$v_{\rm eq}$&$n_{\rm eq}$&$v_{-}$&$n_{-}$&$E_{\rm j}$&$E_{\rm j,rel}$\\
&&[yr]&[GHz]&[$\mu$Jy]&[km/s]&[$\rm cm^{-3}$]&[km/s]&[$\rm cm^{-3}$]&[km/s]&[$\rm cm^{-3}$]&[km/s]&[$\rm cm^{-3}$]&[erg]&[erg]\\
\hline
RXJ1624+7554 (1)&0.06&21.7&3.0&51.0&$120$&$6.5\times10^{8}$&$8000$&$1.7\times10^{2}$&$1200$&$2.3\times10^{6}$&$9000$&$1.4\times10^{3}$&$6.0\times10^{50}$&$3.9\times10^{55}$\\
RXJ1242-1119 (1)&0.046&19.9&3.0&54.0&$100$&$9.2\times10^{8}$&$8500$&$1.1\times10^{2}$&$1000$&$3.3\times10^{6}$&$9400$&$1.1\times10^{3}$&$3.6\times10^{50}$&$3.3\times10^{55}$\\
SDSSJ1323+48 (1)&0.08&8.5&3.0&102.0&$540$&$2.4\times10^{7}$&$8800$&$1.1\times10^{3}$&$5300$&$8.5\times10^{4}$&$9600$&$1.0\times10^{4}$&$6.5\times10^{50}$&$6.1\times10^{54}$\\
SDSSJ1311-01 (1)&0.156&8.3&3.0&57.0&$790$&$1.1\times10^{7}$&$8200$&$2.5\times10^{3}$&$7700$&$4.0\times10^{4}$&$9100$&$2.2\times10^{4}$&$1.4\times10^{51}$&$5.8\times10^{54}$\\
NGC5905 (1,2)&0.012&6.3&8.46&90.0&$41$&$6.8\times10^{10}$&$9900$&$1.9\times10^{2}$&$409$&$2.4\times10^{8}$&$12000$&$1.3\times10^{3}$&$3.0\times10^{49}$&$3.4\times10^{54}$\\
&&21.9&3.0&200.0&$49$&$5.1\times10^{9}$&$9200$&$3.4\times10^{1}$&$480$&$1.8\times10^{7}$&$10000$&$3.2\times10^{2}$&$1.2\times10^{50}$&$4.0\times10^{55}$\\
GALEX-D1-9 (3)&0.326&7.5&5.0&27.0&$710$&$4.4\times10^{7}$&$6900$&$1.2\times10^{4}$&$7000$&$1.5\times10^{5}$&$8300$&$8.1\times10^{4}$&$4.4\times10^{51}$&$4.7\times10^{54}$\\
GALEX-D3-13 (3,4)&0.3698&1.3&1.4&45.0&$21000$&$3.4\times10^{3}$&($9700$)&($5.9\times10^{4}$)&$200000$&$8.3\times10^{0}$&($11000$)&($4.8\times10^{5}$)&$5.2\times10^{50}$&$1.3\times10^{53}$\\
&&8.1&5.0&24.0&$690$&$4.7\times10^{7}$&$6400$&$1.6\times10^{4}$&$6800$&$1.7\times10^{5}$&$8100$&$9.2\times10^{4}$&$5.8\times10^{51}$&$5.4\times10^{54}$\\
D23H-1 (3)&0.186&7.3&4.3&24.0&$490$&$7.1\times10^{7}$&$8500$&$2.4\times10^{3}$&$4800$&$2.5\times10^{5}$&$9400$&$2.2\times10^{4}$&$1.1\times10^{51}$&$4.5\times10^{54}$\\
PTF10iya (3)&0.224&1.7&5.0&24.0&$2200$&$4.7\times10^{6}$&$9700$&$2.2\times10^{4}$&$21000$&$1.7\times10^{4}$&($11000$)&($1.7\times10^{5}$)&$3.5\times10^{50}$&$2.3\times10^{53}$\\
PS1-10jh (3)&0.17&0.83&5.0&45.0&$4600$&$9.6\times10^{5}$&$9900$&$5.9\times10^{4}$&$45000$&$3.4\times10^{3}$&($12000$)&($3.7\times10^{5}$)&$1.5\times10^{50}$&$5.8\times10^{52}$\\
SDSS-TDE1 (3)&0.136&5.4&5.0&30.0&$470$&$1.0\times10^{8}$&$9200$&$2.3\times10^{3}$&$4600$&$3.7\times10^{5}$&$10000$&$2.2\times10^{4}$&$5.5\times10^{50}$&$2.5\times10^{54}$\\
SDSS-TDE2 (3,5)&0.252&0.14&8.4&255.0&$53000$&$1.3\times10^{4}$&($10000$)&($5.5\times10^{6}$)&$\sim300000$&$2.4\times10^{1}$&($13000$)&($2.8\times10^{7}$)&$1.4\times10^{50}$&$1.6\times10^{51}$\\
&&4.4&5.0&36.0&$1100$&$1.7\times10^{7}$&$8600$&$1.1\times10^{4}$&$11000$&$5.9\times10^{4}$&($9400$)&($9.8\times10^{4}$)&$1.8\times10^{51}$&$1.6\times10^{54}$\\
SDSSJ1201+30 (6)&0.146&1.3&1.4&201.0&$19000$&$3.3\times10^{3}$&($9800$)&($3.4\times10^{4}$)&$170000$&$8.3\times10^{0}$&($11000$)&($2.5\times10^{5}$)&$2.8\times10^{50}$&$1.4\times10^{53}$\\
PTF09axc (7)&0.1146&5.0&6.1&150.0&$760$&$4.3\times10^{7}$&$8600$&$6.9\times10^{3}$&$7400$&$1.5\times10^{5}$&$9400$&$6.4\times10^{4}$&$1.5\times10^{51}$&$2.1\times10^{54}$\\
PS1-11af (8)&0.4046&0.24&4.9&51.0&$36000$&$1.5\times10^{4}$&($9900$)&($1.5\times10^{6}$)&$\sim300000$&$3.0\times10^{1}$&($13000$)&($8.6\times10^{6}$)&$1.7\times10^{50}$&$5.0\times10^{51}$\\
&&1.0&5.876&30.0&$5600$&$9.8\times10^{5}$&$9600$&$1.4\times10^{5}$&$55000$&$3.5\times10^{3}$&($11000$)&($1.2\times10^{6}$)&$7.6\times10^{50}$&$8.8\times10^{52}$\\
&&2.4&5.876&45.0&$2900$&$3.4\times10^{6}$&$8500$&$6.7\times10^{4}$&$28000$&$1.2\times10^{4}$&($9400$)&($6.2\times10^{5}$)&$2.9\times10^{51}$&$4.9\times10^{53}$\\
PS16dtm (9)&0.0804&0.11&6.0&23.0&$10000$&$3.8\times10^{5}$&$10000$&$4.1\times10^{5}$&$98000$&$1.2\times10^{3}$&($15000$)&($1.1\times10^{6}$)&$2.7\times10^{48}$&$1.1\times10^{51}$\\
&&0.36&6.0&25.0&$3300$&$3.5\times10^{6}$&$10000$&$6.7\times10^{4}$&$33000$&$1.2\times10^{4}$&($14000$)&($2.5\times10^{5}$)&$1.0\times10^{49}$&$1.1\times10^{52}$\\
iPTF16fnl (10)&0.0163&0.0063&6.1&34.0&$49000$&$2.8\times10^{4}$&($10000$)&($8.4\times10^{6}$)&$\sim300000$&$5.2\times10^{1}$&($19000$)&($1.1\times10^{7}$)&$9.9\times10^{45}$&$3.3\times10^{48}$\\
&&0.0082&15.0&117.0&$27000$&$4.1\times10^{5}$&($10000$)&($1.5\times10^{7}$)&$250000$&$9.3\times10^{2}$&($18000$)&($2.3\times10^{7}$)&$5.1\times10^{46}$&$5.7\times10^{48}$\\
&&0.019&15.0&117.0&$12000$&$2.3\times10^{6}$&($10000$)&($3.9\times10^{6}$)&$110000$&$6.6\times10^{3}$&($18000$)&($6.9\times10^{6}$)&$1.4\times10^{47}$&$3.1\times10^{49}$\\
&&0.052&15.0&117.0&$4300$&$1.7\times10^{7}$&$10000$&$7.9\times10^{5}$&$42000$&$5.9\times10^{4}$&($17000$)&($1.7\times10^{6}$)&$4.9\times10^{47}$&$2.3\times10^{50}$\\
&&0.15&15.0&75.0&$1200$&$2.3\times10^{8}$&$10000$&$1.2\times10^{5}$&$12000$&$8.0\times10^{5}$&$16000$&$3.0\times10^{5}$&$1.1\times10^{48}$&$1.8\times10^{51}$\\
iPTF15af (11)&0.0790&0.044&6.1&84.0&$46000$&$1.4\times10^{4}$&($10000$)&($3.6\times10^{6}$)&$\sim300000$&$2.7\times10^{1}$&($16000$)&($8.9\times10^{6}$)&$2.0\times10^{48}$&$1.7\times10^{50}$\\
AT2018zr (12)&0.071&0.071&16.0&120.0&$12000$&$1.4\times10^{6}$&($10000$)&($2.6\times10^{6}$)&$110000$&$4.2\times10^{3}$&($15000$)&($8.0\times10^{6}$)&$6.3\times10^{48}$&$4.2\times10^{50}$\\
&&0.074&10.0&27.0&$8900$&$1.4\times10^{6}$&$10000$&$9.2\times10^{5}$&$86000$&$4.4\times10^{3}$&($16000$)&($2.4\times10^{6}$)&$1.9\times10^{48}$&$4.7\times10^{50}$\\
&&0.15&10.0&37.5&$5000$&$4.0\times10^{6}$&$10000$&$3.4\times10^{5}$&$49000$&$1.4\times10^{4}$&($15000$)&($1.1\times10^{6}$)&$5.9\times10^{48}$&$2.0\times10^{51}$\\
AT2018fyk (13,14)&0.059&0.032&18.95&38.0&$11000$&$3.3\times10^{6}$&($10000$)&($4.4\times10^{6}$)&$100000$&$9.9\times10^{3}$&($16000$)&($9.8\times10^{6}$)&$8.9\times10^{47}$&$8.6\times10^{49}$\\
&&0.11&18.95&74.0&$4500$&$1.6\times10^{7}$&$10000$&$9.3\times10^{5}$&$44000$&$5.8\times10^{4}$&($15000$)&($2.9\times10^{6}$)&$6.1\times10^{48}$&$9.3\times10^{50}$\\
&&0.21&18.95&53.0&$2000$&$9.3\times10^{7}$&$10000$&$2.7\times10^{5}$&$19000$&$3.3\times10^{5}$&($14000$)&($9.6\times10^{5}$)&$1.1\times10^{49}$&$3.6\times10^{51}$\\
&&0.62&7.25&104.0&$2400$&$8.1\times10^{6}$&$10000$&$4.6\times10^{4}$&$23000$&$2.9\times10^{4}$&($13000$)&($2.2\times10^{5}$)&$3.9\times10^{49}$&$3.2\times10^{52}$\\
&&0.76&7.25&27.0&$1000$&$6.0\times10^{7}$&$10000$&$1.6\times10^{4}$&$9900$&$2.1\times10^{5}$&$13000$&$7.1\times10^{4}$&$1.6\times10^{49}$&$4.9\times10^{52}$\\
&&1.6&7.25&46.0&$620$&$1.4\times10^{8}$&$9900$&$6.6\times10^{3}$&$6100$&$5.0\times10^{5}$&$12000$&$3.9\times10^{4}$&$5.5\times10^{49}$&$2.1\times10^{53}$\\
AT2017eqx (15)&0.1089&0.10&6.0&27.0&$16000$&$1.4\times10^{5}$&($10000$)&($7.5\times10^{5}$)&$150000$&$3.6\times10^{2}$&($15000$)&($2.2\times10^{6}$)&$4.2\times10^{48}$&$8.9\times10^{50}$\\
&&0.21&6.0&26.0&$7700$&$5.9\times10^{5}$&$10000$&$2.4\times10^{5}$&$75000$&$2.0\times10^{3}$&($14000$)&($8.3\times10^{5}$)&$9.9\times10^{48}$&$3.7\times10^{51}$\\
XMMSL2&0.029&0.066&6.0&28.0&$7400$&$9.9\times10^{5}$&$10000$&$3.3\times10^{5}$&$72000$&$3.4\times10^{3}$&($17000$)&($6.9\times10^{5}$)&$3.1\times10^{47}$&$3.7\times10^{50}$\\
~~~J1446+68 (16)&&0.51&6.0&18.0&$780$&$9.6\times10^{7}$&$10000$&$1.0\times10^{4}$&$7700$&$3.4\times10^{5}$&$15000$&$3.2\times10^{4}$&$1.9\times10^{48}$&$2.1\times10^{52}$\\
ASASSN-18pg (17)&0.0174&0.026&18.95&50.0&$4900$&$2.4\times10^{7}$&$10000$&$1.8\times10^{6}$&$48000$&$8.7\times10^{4}$&($17000$)&($3.2\times10^{6}$)&$1.4\times10^{47}$&$5.6\times10^{49}$\\
&&0.073&18.95&43.0&$1600$&$2.3\times10^{8}$&$10000$&$3.2\times10^{5}$&$16000$&$8.2\times10^{5}$&$17000$&$6.9\times10^{5}$&$4.1\times10^{47}$&$4.5\times10^{50}$\\
\hline
\multicolumn{15}{l}{{\bf Ref.} (1) \citealt{Bower+2013}, (2) \citealt{Komossa2002}, (3) \citealt{vanVelzen+2013}, (4) \citealt{Bower2011}, (5) \citealt{vanVelzen+2011},  (6) \citealt{Saxton+2012}, (7) \citealt{Arcavi+2014}}\\
\multicolumn{15}{l}{(8) \citealt{Chornock+2014}, (9) \citealt{Blanchard+2017}, (10) \citealt{Blagorodnova+2017}, (11) \citealt{Blagorodnova+2019}, (12) \citealt{vanVelzen+2019}, (13) \citealt{Wevers+2019}}\\
\multicolumn{15}{l}{(14) \citealt{Wevers+2021}, (15) \citealt{Nicholl+2019}, (16) \citealt{Saxton+2019}, (17) \citealt{Holoien+2020}}
\end{tabular}
\end{table}
\end{landscape}

\begin{landscape}
\begin{table}
\caption{Required conditions to produce the observed radio TDEs for the cases of isotropic disk wind and unbound debris.
For events without measured spectral index or electron power-law index $p$, we adopt $p=2.5$.
When the spectral peak is available, we show only the minimal velocity $v_{\rm eq}$ and corresponding density $n_{\rm eq}$ comparable to those obtained in the equipartition analysis.
For the last observation of CNSS J0019+00 without the spectral peak, we adopt the initial wind velocity of $v_{\rm in}=8000\,\rm km\,s^{-1}$.}
\label{table detection}
\begin{tabular}{llllrr|rrrr|rrrr}
\hline
Event (Ref)&Redshift&Index&Time&Freq.&Flux&\multicolumn{4}{c|}{Disk wind ($\Omega=4\pi$)}&\multicolumn{4}{c}{Unbound debris ($\Omega=0.1$)}\\
&$z$&$p$&$t$&$\nu$&$F_{\nu}$&$v_{\rm eq}$&$n_{\rm eq}$&$v_{-}$&$n_{-}$&$v_{\rm eq}$&$n_{\rm eq}$&$v_{-}$&$n_{-}$\\
&&&[yr]&[GHz]&[$\mu$Jy]&[$\rm km/s$]&[$\rm cm^{-3}$]&[$\rm km/s$]&[$\rm cm^{-3}$]&[$\rm km/s$]&[$\rm cm^{-3}$]&[$\rm km/s$]&[$\rm cm^{-3}$]\\
\hline
IGR J12580+0134 (1)&0.004&(2.5)&1.0&6.0&414000&$7100$&$2.9\times10^{5}$&$9700$&$9.2\times10^{4}$&$69000$&$9.9\times10^{2}$&($11000$)&($7.2\times10^{5}$)\\
&&&1.145&6.0&355000&$5700$&$4.6\times10^{5}$&$9700$&$6.8\times10^{4}$&$56000$&$1.6\times10^{3}$&($11000$)&($5.5\times10^{5}$)\\
&&&1.529&6.0&209000&$3300$&$1.5\times10^{6}$&$9700$&$3.3\times10^{4}$&$33000$&$5.4\times10^{3}$&($11000$)&($2.7\times10^{5}$)\\
\hline
XMMSL1 J0740-85 (2)&0.0173&3.0&1.6&9.0&380&$590$&$3.6\times10^{8}$&$9800$&$1.9\times10^{4}$&$5800$&$1.3\times10^{6}$&$11000$&$1.2\times10^{5}$\\
&&&1.668&18.0&130&$170$&$2.1\times10^{10}$&$9800$&$1.5\times10^{4}$&$1700$&$7.8\times10^{7}$&$12000$&$9.4\times10^{4}$\\
&&&2.107&9.0&250&$370$&$9.8\times10^{8}$&$9700$&$1.1\times10^{4}$&$3700$&$3.6\times10^{6}$&$11000$&$7.1\times10^{4}$\\
&&&2.397&9.0&230&$310$&$1.4\times10^{9}$&$9700$&$8.7\times10^{3}$&$3100$&$5.2\times10^{6}$&$11000$&$5.9\times10^{4}$\\
\hline
CNSS J0019+00 (3)&0.018&3.3&1.7$^a$&3.9&8080&$7100$&$3.1\times10^{5}$&-&-&$70000$&$1.1\times10^{3}$&-&-\\
&&&2.0$^a$&3.1&7410&$7000$&$2.1\times10^{5}$&-&-&$69000$&$7.3\times10^{2}$&-&-\\
&&&2.7$^a$&1.9&4950&$6900$&$8.9\times10^{4}$&-&-&$68000$&$3.2\times10^{2}$&-&-\\
&&&4.2&1.9&1350&$2500$&$8.4\times10^{5}$&$6700$&$2.7\times10^{4}$&$25000$&$3.2\times10^{3}$&($9700$)&($8.1\times10^{4}$)\\
\hline
ASASSN-14li (4,5)&0.0206&3.0&0.39$^a$&8.2&1760&$6500$&$1.6\times10^{6}$&-&-&$64000$&$6.0\times10^{3}$&-&-\\
&&&0.57$^a$&4.4&1230&$7100$&$4.2\times10^{5}$&-&-&$70000$&$1.5\times10^{3}$&-&-\\
&&&0.67$^a$&4.0&1140&$6300$&$4.6\times10^{5}$&-&-&$63000$&$1.7\times10^{3}$&-&-\\
&&&0.83$^a$&2.5&940&$7300$&$1.4\times10^{5}$&-&-&$71000$&$4.9\times10^{2}$&-&-\\
&&&1.0$^a$&1.9&620&$6400$&$1.1\times10^{5}$&-&-&$63000$&$4.2\times10^{2}$&-&-\\
\hline
AT2019dsg (6,7)&0.051&2.7&0.15$^a$&16.2&560&$9600$&$2.1\times10^{6}$&-&-&$92000$&$6.4\times10^{3}$&-&-\\
&&&0.23$^a$&10.5&740&$11000$&$5.9\times10^{5}$&-&-&$110000$&$1.7\times10^{3}$&-&-\\
&&&0.44$^a$&7.9&1110&$9300$&$4.7\times10^{5}$&-&-&$89000$&$1.4\times10^{3}$&-&-\\
&&&0.82$^a$&3.8&890&$9400$&$1.1\times10^{5}$&-&-&$90000$&$3.4\times10^{2}$&-&-\\
&&&1.5$^a$&1.8&380&$7200$&$5.0\times10^{4}$&-&-&$70000$&$1.7\times10^{2}$&-&-\\
\hline
AT2020zso (7)&0.061&(2.5)&0.11&15.0&22&$3300$&$2.5\times10^{7}$&$10000$&$4.5\times10^{5}$&$32000$&$9.0\times10^{4}$&($15000$)&($1.2\times10^{6}$)\\
\hline
AT2020vwl (8)&0.035&(2.5)&0.38&9.0&552&$4200$&$3.3\times10^{6}$&$10000$&$1.5\times10^{5}$&$41000$&$1.2\times10^{4}$&($13000$)&($6.7\times10^{5}$)\\
\hline
\multicolumn{14}{l}{$^a$ Spectral peak.}\\
\multicolumn{14}{l}{{\bf Ref.} (1) \citealt{Irwin+2015}, (2) \citealt{Alexander+2017}, (3) \citealt{Anderson+2020}, (4) \citealt{Alexander+2016}, (5) \citealt{vanVelzen+2016},  (6) \citealt{Stein+2021}}\\
\multicolumn{14}{l}{(7) \citealt{Cendes+2021b}, (8) \citealt{Alexander+2021}, (9) \citealt{Goodwin+2021}}\\
\end{tabular}
\end{table}
\end{landscape}

\end{document}